\documentclass[twocolumn,floats,amsmath,amssymb,preprintnumbers,superscriptaddress,floatfix]{revtex4-1}

\usepackage{graphicx}
\usepackage{bm}
\usepackage{color}
\usepackage{hyperref}
\usepackage{ulem}
\usepackage{amsmath}
\usepackage{amssymb}
\extrafloats{100}

\def\Eq{eq}
\def\Eqs{eqs}
\def\Figure{Figure}
\def\Figures{Figures}
\def\Fig{Figure}
\def\Figs{Figures}

\begin{document}

\title{Cosolute partitioning in polymer networks: Effects of flexibility and volume transitions}

\author{Won Kyu Kim}
\email{wonkyu.kim@helmholtz-berlin.de}
\affiliation{ Institut f\"ur Weiche Materie und Funktionale Materialien, Helmholtz-Zentrum Berlin, Hahn-Meitner-Platz~1,14109 Berlin, Germany}

\author{Arturo Moncho-Jord\'{a}}
\affiliation{ Departamento de F\'{i}sica Aplicada, Facultad de Ciencias, Universidad de Granada, Avenida Fuente Nueva S/N, 18071 Granada, Spain}
\affiliation{ Instituto Carlos I de F\'{i}sica Te\'{o}rica y Computacional, Facultad de Ciencias, Universidad de Granada, Avenida Fuente Nueva S/N, 18071 Granada, Spain}

\author{Rafael Roa}
\affiliation{ Institut f\"ur Weiche Materie und Funktionale Materialien, Helmholtz-Zentrum Berlin, Hahn-Meitner-Platz~1,14109 Berlin, Germany}

\author{Matej Kandu\v{c}}
\affiliation{ Institut f\"ur Weiche Materie und Funktionale Materialien, Helmholtz-Zentrum Berlin, Hahn-Meitner-Platz~1,14109 Berlin, Germany}

\author{Joachim Dzubiella}
\email{joachim.dzubiella@helmholtz-berlin.de}
\affiliation{ Institut f\"ur Weiche Materie und Funktionale Materialien, Helmholtz-Zentrum Berlin, Hahn-Meitner-Platz~1,14109 Berlin, Germany}
\affiliation{ Institut f{\"u}r Physik, Humboldt-Universit{\"a}t zu Berlin, Newtonstr.~15, 12489 Berlin, Germany}


\begin{abstract}
	We study the partitioning of cosolute particles in a thin film of a semi-flexible polymer network by a combination of 
	coarse-grained (implicit-solvent) stochastic dynamics simulations and mean-field theory. 
	We focus on a wide range of solvent qualities and cosolute--network interactions for selected polymer flexibilities. 
	Our investigated ensemble (isothermal-isobaric) allows the network to undergo a volume transition from extended to collapsed state while the cosolutes can distribute in bulk and network, correspondingly.    
	We find a rich topology of equilibrium states of the network and transitions between them, qualitatively depending on solvent quality, polymer flexibility, and cosolute--network interactions. In particular, we find a novel `cosolute-induced' collapsed state, where strongly attractive cosolutes bridge network monomers albeit the latter interact mutually repulsive. Finally, the cosolutes' global partitioning `landscape', computed  as a function of solvent quality and cosolute--network interactions, exhibits very different topologies depending on polymer flexibility.  The simulation results are supported by theoretical predictions obtained with a two-component mean-field approximation for the Helmholtz free energy that considers the chain elasticity and the particle interactions in terms of a virial expansion. Our findings have implications on the interpretation of transport processes and permeability  in hydrogel films, as realized in filtration or macromolecular carrier systems.
\end{abstract}

\maketitle


\section{Introduction}

Polymer networks such as microgels or hydrogels are integral components of modern soft functional material design. 
Important applications revolve around filtration, macromolecular carrier particles, functional coatings,  sensors,  
and nanoreactors.~\cite{Peppas, Hamidi, Tokarev, stuart2010emerging, Lu2011,welsch2010microgels,lu2006thermosensitive,lu2007thermosensitive, Menne2014,Plamper2017}  In all these systems the uptake, storage,  and release of active agents and cosolutes (e.g., reactants,  pollutants, ligands, drugs) are decisive processes, which have to be controlled  and programmed to the material. The gels in these applications often form film-like layers \cite{wu1994studies,zhou1996situ}, shells, and membranes that modulate the permeation \cite{lobaskin2012interactions,petrache2006swelling} and separation of the cosolutes by their wide range of controllable physicochemical behavior. 
 A fundamental property describing the permeation and sorption of cosolutes to polymer networks on 
the most global level is the partitioning 
\begin{eqnarray} 
\mathcal{K} = c_{c}^\text{in} / c_{c}^\text{out},
\end{eqnarray}
which is a ratio of cosolute concentrations inside and outside (in the bulk reservoir) of the gel \cite{Leo1971}. 
Equilibrium partitioning is one of the two key ingredients to the {\it permeability} of hydrogels that eventually defines the 
degree of mass transport through the networks \cite{gehrke,Stefano2015,Rafa2017}.  

One notable physical property that distinguishes gels from ordinary polymeric solutions is the additional inter-connectivity that cross-links individual polymer molecules. This higher level complexity yields an intriguing nature of the material that is liquid-like and also solid-like \cite{shibayama1993volume}, which renders a well known transition phenomenon: Most hydrogels undergo a sharp reversible structural transition from swollen states to collapsed states~\cite{khokhlov1980swelling,erman1986critical,khokhlov1993conformational,barenbrug1995highly,heskins1968solution,duvsek1968transition, Habicht2015}.  This volume transition is a reversible function of the temperature \cite{zhou1996situ,wu1997volume}, or more general, solvent quality, but can also be triggered  for some polymers by several environmental stimuli,  such as pH, salinity \cite{stuart2010emerging}, and, importantly in our context,  the addition of co-solvents or co-solutes \cite{conon, zhang, Heyda2013, Heyda2014}. The responsiveness close to the volume transition enables a facilitated control of the gel structure and permeability, eventually bearing a number of applications in the realm of so-called stimuli-responsive, `smart' polymer materials \cite{zhao2001nonergodic,stuart2010emerging}.

Hence, many relevant and intertwined phenomena come into play for the partitioning of cosolutes into gels, such as network topology and heterogeneity, size exclusion, as well as attractive interactions such as van der Waals, hydrophobic and electrostatic interactions, strongly depending on the chemical nature of polymers, solvents, and cosolutes \cite{moncho2013effective,moncho2013effective2,moncho2014ion,doi:10.1021/acs.langmuir.7b00356,Matej2017}. 
In responsive polymer materials, as referenced above, the presence and adsorption of cosolutes can lead to swelling  or shrinking or even 
induce a volume transition of the network. Hence, in general the detailed mechanisms of partitioning in polymer networks can be expected to be utterly complex as they result from a subtle interplay of non-bonded interactions, network constraints, comformations, and their transitions on multiple scales. Therefore, little is known yet about the quantitative behavior of cosolute partitioning that can be raised during the swelling--collapse transition of the gels in contact with a reservoir of the cosolutes, despite the elaborate development of theories \cite{shibayama1993volume,Tanaka1978,tanaka1977critical,Mann2007} and simulations\cite{Duering1994,Escobedo1997,Escobedo1999,Aydt2000,Lu2002,Schneider2002,Schneider2003,Schneider2004,Yan2003,Edgecombe2007,Binder2008,Jha2011,Jiao2012,Quesada-Perez2012,Vagias2013,Kosovan2015,Kobayashi2016,Schmid2016} for pure gel systems (no cosolutes included) over the past decades. Noteworthy exceptions are recent simulation works that study cosolute permeation of hollow nanogels~\cite{Schmid2016} or 
the non-specific effects of charged solutes (mostly salt) and their electrostatic couplings in polyelectrolyte network gels \cite{Schneider2002,Schneider2003,Schneider2004,Yan2003,Edgecombe2007,Jha2011,Jiao2012,Quesada-Perez2012,Vagias2013,Kosovan2015,adroher2015role,Kobayashi2016,adroher2017competition}. 
	
In this work, we build a  minimalistic model system to study the partitioning of cosolute particles in a thin film of a semi-flexible polymer network 
by a combination of coarse-grained stochastic dynamics simulations and mean-field theory. We perform a systematic study focusing on effects of varying solvent qualities (from good to bad), specific cosolute--network interactions (from repulsive to highly attractive) as well as polymer flexibilities.
We work in an isothermal-isobaric ensemble in order to include also the effects of network swelling, shrinking, and the volume transition, coupled to the equilibrium partitioning of the cosolutes from a reservoir. To start as simple as possible, we construct a coarse-grained gel consisting of a regular cubic network \cite{Aydt2000,Lu2002,pnetz1997,Erbas2015,Erbas2016,Li2016} of flexible or semi-flexible chains in the presence of cosolutes which are simply of monomer size, and we employ inter- and intra-particle interactions in terms of the generic  Lennard-Jones (LJ) pair potentials, which can be easily tuned from repulsion to high attraction.   Despite the minimalistic features of our model, we find very rich structural and partitioning topologies, which we summarize in state--diagrams in the solvent-quality--interaction space. A large influence of polymer flexibility, in terms of the gel volume response, transition order and sharpness, on the partitioning is observed.  Our mean-field theory is fully consistent with the simulations and provides a theoretical interpretation of the complex coupled partitioning and network volume (transition) behavior.  

\section{Methods}\label{sec:methods}
	
\subsection{Simulation model}

In our simulation model we consider a film of a cubic polymer network in implicit solvent and explicit cosolutes, where monomers, linkers, and cosolutes are represented by simple beads. 
 For the polymers we employ a bead--spring model with the intra-chain Hamiltonian (in units of thermal energy, $k_\text{B} T \equiv 1/\beta$) as
\begin{eqnarray}\label{eq:poten0}
U^{\text{B}} _i= k_s (r_{i, i+1} - \sigma)^2 + k_b (\theta_{i-1,i,i+1} - \theta_0)^2.
\end{eqnarray}
The first term is the harmonic bond potential with respect to the segmental bond distance $r_{i,i+1} = | {\bf{r}}_{i+1} - {\bf{r}}_{i} |$, with the spring constant $k_s = 80 ~k_\text{B} T/\sigma^2$ and the equilibrium bond length $\sigma$. The latter sets the length scale of our system.  The second term is a harmonic bending potential with respect to the internal angle $\theta_{i-1,i,i+1}$ of two adjacent segmental vectors ${{\bf{r}}_{i+1}-\bf{r}}_{i}$ and ${{\bf{r}}_{i}-\bf{r}}_{i-1}$, with a rigidity constant $k_b$. This potential we use to tune the network flexibility. Here, we choose the equilibrium angle $\theta_0 = 180^\circ$ \cite{Schneider2003,Edgecombe2007} to avoid additional complexity of introducing torsional potentials. 
With $k_b$ we can interpolate between flexible chains, $k_b = 0$,  to semi-flexible ones with $k_b$ in the range of several $k_\text{B} T$, and finally very stiff networks for $k_b \gg k_\text{B} T$. For the cross-links we define cross-linking beads that are bonded analogously as described in \Eq~\eqref{eq:poten0} but  in three perpendicular directions in order to establish a regular cubic network in 3D. Hence, a cross-linking bead has six neighbors, unless it is  a surface bead in the thin film having only five neighbors. We do not consider an angular potential involving the cross-linkers~\cite{Schneider2003}.

In addition to the intra-bead potentials, all network beads and cosolute particles are subject to the Lennard-Jones (LJ) potential:
\begin{equation}\label{eq:LJ}
U^{\text{LJ}}_{ij} = 4 \epsilon_{ij} \left[ \left( \frac{\sigma_{ij}}{r_{ij}} \right)^{12} - \left( \frac{\sigma_{ij}}{r_{ij}} \right)^{6} \right],
\end{equation}
where $\sigma_{ij}$ is the interaction size, and the parameter $\epsilon_{ij}$ (in units of $k_\text{B} T$) is the interaction energy parameter for $i,j=n,c$, that is, network and cosolute particles, respectively. For network monomers, $i,j=n$, we also refer to it $\epsilon_{nn}$ as the solvent quality parameter as  it tunes from a perfect ideal polymer behavior ($\epsilon_{nn}=0$) to a good solvent ($0< \beta \epsilon_{nn}<0.3$), passing the $\theta$-solvent ($\beta \epsilon_{nn} = 0.3$, where the second virial coefficient for the LJ interaction vanishes) to a bad solvent ($\beta \epsilon_{nn}> 0.3$)~\cite{Heyda2013,Schmid2016}. 
The parameter $\epsilon_{nc}$ refers to the cosolute coupling to the polymers, which also will be tuned from the ideal behavior to very attractive interactions.  Furthermore, we fix the cosolute--cosolute interaction to a generic value $\epsilon_{cc} = 0.1 ~k_\text{B} T$, that ensures simple repulsion between the cosolute particles and avoids unwanted aggregation between them. For simplicity we keep $\sigma_{ij}\equiv \sigma$ for all the beads. 


The thin film of a cubic network is initially located in the middle of the simulation box in a stretched form, periodically repeated in $x$- and $y$-directions, while in $z$-direction it is finite and in touch with a finite reservoir (see \Fig~1 and \Fig~S1 in SI).
 The network film consists of 6 layers (in the $xy$ plane) of 5$\times$5 cross-linker beads, where every cross-linker is connected linearly by 4 beads so that the inter-cross-linker length (contour length) is $l=5\sigma$. With that, there are 425 chains and 150 cross-linkers with total network particle number $N_n = 1850$, and the fraction of the cross-linker beads yields $8.1\%$. 
We perform simulations spanning a wide range of the rigidity constant, $k_b$,
and the interaction parameters, $\epsilon_{ij}$, and investigate how the system (with or without cosolutes) is affected by these parameters.  In the systems including cosolutes we add $N_{c} = 576$ cosolute particles,   amounting to an initial concentration of $c_{c}\approx 0.012 \sigma^{-3}$. The resulting packing fraction is $\phi_{c} = \pi \sigma^3 c_{c} / 6 \approx 0.0064$. We thus work in the high dilution limit of cosolutes in the reservoirs.  

\subsection{Simulation protocol}

We employ the LAMMPS software \cite{Plimpton1995} with the stochastic Langevin integrator in an anisotropic $NpT$ ensemble. The iteration time step $\delta \tau = 10^{-3} \tau$ defines the time unit $\tau = \sqrt{ {m \sigma^2} / {k_\text{B} T} }$. The friction coefficient $\gamma$ is chosen to have the momentum relaxation time $\tau_{\gamma} = m/\gamma = \tau$, so that the free cosolute motion becomes diffusive after $10^3$ time steps.
To maintain fixed pressure we use the Berendsen anisotropic barostat~\cite{berendsen1984molecular}.   The pressure relaxation time is also chosen to be $\tau_{p} = \tau$.   The simulation box is initially set to have the lengths $L_x = L_y = 25 \sigma$ and the longitudinal length $L_z = 75 \sigma$, with periodic boundary conditions in all three Cartesian directions. The long box length $L_z$ is kept fixed, while $L_x $ and $L_y$ vary according to $NpT$ ensembles with given particle number $N = N_n + N_{c} = 2426$, pressure $p_x = p_y = p$, and temperature $T$. In that way, the hydrogel is allowed to self-consistently sample its equilibrium volume and aspect ratio, given the prescribed interactions. As the hydrogel is connected to a (finite) reservoir of cosolutes, the cosolutes can always equilibrate their partitioning between the reservoir and the gel.  The value of the lateral pressure is chosen to be $p = 6.5853 \times 10^{-4} k_\text{B} T/\sigma^3 \approx 1$ bar.  
The simulations are typically run up to $4 \times 10^6 \delta \tau = 4000 \tau$ and we analyze the data in the last 20\% of the production runs. 

\begin{figure}
\centering
\includegraphics[width = 0.4\textwidth]{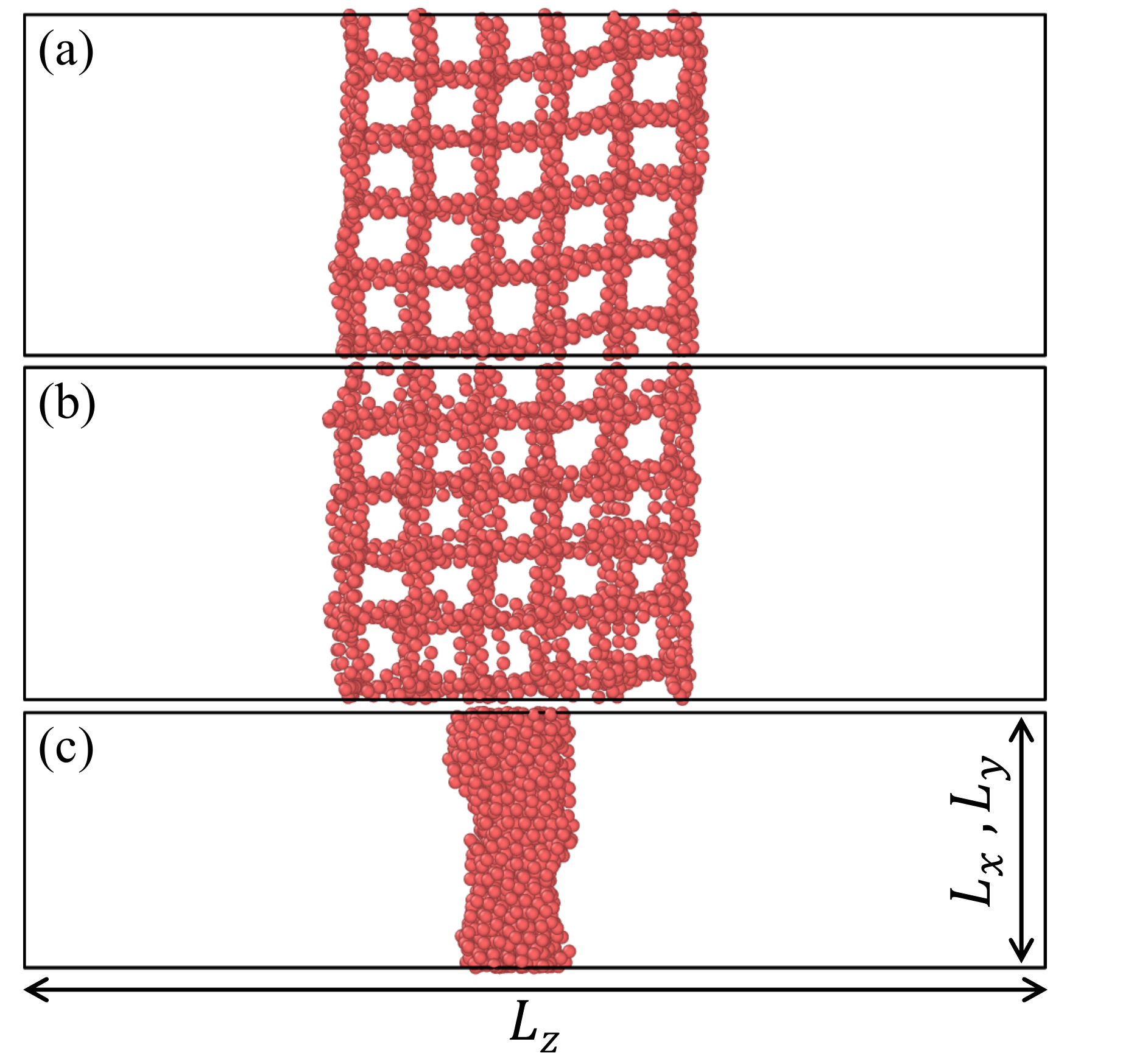}
\caption{Snapshots of the simulations for the cubic network film at equilibrium without cosolutes. The rectangular frames depict the actual simulation box of lengths $L_x, L_y$ and $L_z$, shown for the following cases: 
(a) Stiff network with the rigidity constant $k_b = 10 ~k_\text{B} T$ and the intra-network interaction parameter $\epsilon_{nn} = 0.1 ~k_\text{B} T$ (i.e., corresponding to a good solvent).
(b) Semi-flexible network with $k_b = 3 ~k_\text{B} T$ and $\epsilon_{nn} = 0.9 ~k_\text{B} T$, where the system is close to the network collapse transition point.
(c) Collapsed semi-flexible network with $k_b = 3 ~k_\text{B} T$ and $\epsilon_{nn} = 2 ~k_\text{B} T$ (i.e., corresponding to a  bad solvent).
}\label{fig:fig1}
\end{figure}

As a first illustration, \Fig~\ref{fig:fig1} shows snapshots of three representative equilibrium states of the network film conformations.  When the network chains are sufficiently stiff and the network monomers are essentially repulsive ($\epsilon_{nn} = 0.1 ~k_\text{B} T$), as shown in \Fig~\ref{fig:fig1}(a), the network structure is in a swollen state, i.e., in a good solvent.  On the other hand, the swollen state is disfavored by large attractions between the network monomers, that eventually result in a collapsed state as shown in \Fig~\ref{fig:fig1}(c), i.e., in a bad solvent. \Figure~\ref{fig:fig1}(b) shows a semi-flexible network system close to its collapse transition point.

\subsection{Models of flexible and semi-flexible networks}

	In most of the paper, we will focus on two network types that we coin `flexible' and `semi-flexible' and shall relate (as far as our minimal model allows) to representative polymers that are widely used in experiments. The first polymer could be based on a simple low-molecular mass monomer like ethylene glycol.  It is known that the persistence length $l_p$ of a poly-ethylene glycol (PEG) chain is $l_p \approx 0.38 ~\text{nm}$ \cite{Kienberger2000,Lee2008}. In our coarse-grained chain model this would correspond roughly to a  bead (monomer) size $\sigma$. Therefore, we focus on the limiting $k_b = 0 $ case for our model flexible network. The second one could be a higher-molecular weight polymer such as poly(N-isopropylacrylamide) (PNIPAM) or poly(methylmethacrylate) with more steric restraints and therefore stiffer conformations, leading to $l_p$ values in the nanometer range~\cite{zhang2000single,ahmed2009uv,kutnyanszky2012there, PMMA}. Given our intra-bond Hamiltonian \Eq~\eqref{eq:poten0}, the bending energy of our model polymers can be  expressed by the persistence length $l_p$ in the limit of small bending via \cite{harris1966polymer, fixman1973polymer, Kierfeld2004}
\begin{eqnarray}\label{eq:persist2}
l_p = \frac{2 k_b \sigma}{k_\text{B} T}.
\end{eqnarray}
The bending rigidity of $k_b = 3 ~k_\text{B} T$, for example, yields the persistence length $l_p \approx 2 ~\text{nm}$, right in the reported range for PNIPAM.  Hence, in the bulk part of our analysis we focus on two cases we define as flexible and semi-flexible networks with chosen rigidity constants $k_b = 0$ and $k_b = 3~k_\text{B} T$, respectively.

\subsection{Two-component (network--cosolute) mean-field theory}
\label{sec:meanfield}

With the purpose of establishing a firm theoretical model that mimics the simulation conditions and is able to explain the cosolute effects on the polymer network states, we develop a mean-field theory for a closed (canonical) system that contains the polymer gel (formed by interconnected and cross-linked monomeric units and cosolutes) and the bulk reservoir (with only cosolutes). Therefore, the total number of cosolutes $N_{c}=N_{c}^{\text{in}}+N_{c}^{\text{out}}$ and the number of monomers in the gel phase $N_{n}$ are both fixed quantities. The total Helmholtz free energy of such system can be written as $F=F_{\rm network}+F_{\rm bulk}$. The model essentially combines standard ideas from elastic polymer and network models\cite{Rubinstein}, extended to include semi-flexibility via the worm-like chain approach~\cite{Blundell2009,Meng2017} and multi-component interactions on the virial expansion level~\cite{Heyda2013}. Within these approximations, the free energy of the two-component network plus cosolute system is split into five different contributions:
\begin{equation}\label{eq:free_energy}
F_{\rm network}=F_{\rm ela}+F_{\rm conf}+F_{\rm ideal}+F_{\rm HS}+F_{\rm int}.
\end{equation}
The first contribution, $F_{\rm ela}$, corresponds to the elastic free energy of the network. For semi-flexible and inextensible polymer chains, it can be approximated using the wormlike chain model as~\cite{Blundell2009,Meng2017}
\begin{eqnarray}
\beta F_{\rm ela}=N_{\rm ch} \left[ \frac{\pi^2 l_p}{2l}\left\{ 1-(R/l)^2 \right\} +\frac{2l}{\pi l_p}\frac{1}{1-(R/l)^2} \right],
\end{eqnarray}
where $R$ is the average end-to-end distance of the polymer chains between two cross-linkers, $l = N_{m} \sigma$ is the contour length ($N_{m}$ is the number of monomers per chain), and $N_{\rm ch}$ is the number of chains of the network. In the limit of full stretching ($R \rightarrow l$) the elastic free energy diverges, as expected for inextensible chains. In the limit of small persistence length and weak stretch ($R\ll l$), this model tends to $\beta F_{\rm ela}\approx \text{const.} +\frac{2R^2}{\pi N_{m}\sigma l_p}$, and for $l_p=4\sigma/(3\pi)$ the  Gaussian flexible chain is recovered. The second contribution in \Eq~\eqref{eq:free_energy}, $F_{\rm conf}$, represents the (entropic) free energy of confinement that arises for highly collapsed states ($R \rightarrow 0$). It can be written as~\cite{Edwards1969} 
\begin{equation}\label{eq:fconf}
\beta F_{\rm conf} = C N_{\rm ch}\frac{\pi^2}{2}\frac{N_{m}\sigma^2}{R^2},
\end{equation}
where $C$ is a phenomenological normalization constant that depends on the network stiffness. This constant takes into account the fact that collapsing a stiff chain always implies a larger free energy than shrinking a flexible (Gaussian) chain. The third contribution is the ideal free energy of the cosolutes,
\begin{equation}
\beta F_{\rm ideal} = c_{c}^\text{in}\left[ \ln(c_{c}^\text{in} \Lambda_{c}^3)-1\right]V_{n},
\end{equation} 
where $\Lambda_{c}$ is the thermal wavelength of the cosolute particles, $V_{n}$ is the volume of the network, and $c_{c}^\text{in} = N_{c}^\text{in} /V_{n}$ is the number density of cosolutes inside the network. 
The fourth contribution is the hard sphere contribution, which takes into account the excluded-volume of the particles and  prevents the interpenetration of monomers and cosolutes in conditions of strong inter-particle attractions. Assuming, just as in the simulations, that monomers and cosolutes have the same size, $\sigma$, this free energy is given by the Carnahan-Starling expression
\begin{equation}
\beta F_{\rm HS} = \frac{4(\phi_n+\phi_{c}^\text{in})-3(\phi_n+\phi_{c}^\text{in})^2}{(1-\phi_n-\phi_{c}^\text{in})^2}(c_{n}+c_{c}^\text{in})V_{n}
\end{equation} 
where $c_{n}=N_{n}/V_{n}$ is the number density of monomers inside the network, $\phi_{n}=\pi c_{n} \sigma^3/6$ its packing fraction, and $\phi_{c}^\text{in}=\pi c_{c}^\text{in} \sigma^3/6$ is the packing fraction of cosolutes inside the network. 

Finally, $F_{\rm int}$ is the interaction free energy between monomers and cosolute particles. Up to second order in the virial expansion of the monomer and cosolute number densities ($c_{n}$ and $c_{c} \equiv c_{c}^\text{in} $, respectively), it is given by
\begin{eqnarray}\label{eq:f_int}
\beta F_{\rm int}=V_{n}\sum_{i,j=n,c}c_i c_j\big(B_2^{ij}-B_{2, \rm HS}^{ij}\big)
\end{eqnarray} 
where $B_2^{ij}$ is the second virial coefficient for LJ interactions. In the expression we substract the second virial coefficient for hard spheres ($B_{2,\rm HS}^{ij}=2\pi \sigma^3/3$), since this contribution has been already included in $F_{\rm HS}$. The second virial coefficients can be calculated numerically for the different combinations of interparticle interaction strengths $\epsilon_{nn}$, $\epsilon_{nc}$, for a fixed intercosolute interaction $\epsilon_{cc}=0.1~k_\text{B} T$.

To obtain the free energy of the bulk phase, $F_{\rm bulk}$, we proceed in the same way and assume the same approximations, but with cosolutes as the only component 
\begin{eqnarray}\label{eq:f_bulk}
\beta F_{\rm bulk}&=& \Bigg[ c_{c}^\text{out}\left( \ln(c_{c}^\text{out} \Lambda_{c}^3)-1\right) +\frac{4\phi_{c}^\text{out}-3(\phi_{c}^\text{out})^2}{(1-\phi_{c}^\text{out})^2}c_{c}^\text{out}  \nonumber \\
&+& (c_{c}^{\rm out})^2\big(B_2^{cc}-B_{2,{\rm HS}}^{cc}\big) \Bigg]V_{\rm bulk}
\end{eqnarray}
where $V_{\rm bulk}$ is the volume of the bulk phase. Note that in this case we use the concentration of cosolutes in the bulk reservoir in the free energy expression, $c_{c}^\text{\rm out}=N_{c}^{\text{out}}/V_{\rm bulk}$, where $N_{c}^{\text{\rm out}}=N_{c}-N_{c}^{\text{in}}$ to impose a constant total number of cosolutes in the whole system.

In order to compare with the simulation results, we assume that the contour length of the chains is $l = N_{m} \sigma = 5\sigma$, and that the polymer network has a cubic structure. The network is composed by 125 cubic cells, each one with a volume $R^3$. Since each cell has 3 chains (with 4 interconnected monomer beads each) and 1 cross-linker monomer, the total volume of the network may be expressed as $V_n=N_{\rm ch}R^3/3$, and the number density of network particles is $c_n=13/R^3$. Following the morphology of the simulation box, we approximate in our theory $V_{\rm bulk} \approx 2V_n$ for simplicity. Moreover, the total number of cosolute particles has been fixed in the theory to be $N_t=576$. Therefore, the resulting total free energy of the system is, for a given temperature, a function of two variables, $F(N_{c}^{\text{in}},V_{n})$. In order to calculate the equilibrium swelling state of the gel network and the concentration of cosolutes absorbed inside we need to minimize the free energy against $N_c^{\text{in}}$ and $V_{n}$ at a constant pressure given by $p = 6.5853 \times 10^{-4} k_\text{B} T/\sigma^3$. This leads to the following non-linear set of algebraic equations:
\begin{eqnarray}\label{eq:eq12}
\left(\frac{\partial F}{\partial N_{c}^{\text{in}}}\right) = 0  \;\;\;\;\;  {\rm and} \;\;\;\;\;
\left(\frac{\partial F}{\partial V_{n}}\right) = -3p
\end{eqnarray}
where the factor of 3 in the last equation comes from the fact that the entire volume of the system is $3 V_{n}$.

We use this model to determine the swelling behavior of the network for two different conditions of chain bending rigidities, namely flexible ($l_p=4\sigma/(3\pi)$, $k_b \approx 0.2~k_\text{B} T$) and semi-flexible ($l_p=6\sigma$, $k_b =3~k_\text{B} T$). 
The phenomenological constant $C$ in \Eq~\eqref{eq:fconf} for the flexible chains is $C=1$, and the resulting swelling curve predicted by the mean-field theory is in very good agreement with the simulation result (for example, see \Fig.~\ref{fig:fig2}) with no free fitting parameters.
For semi-flexible chains, we use a fitted value of $C=3$ at which yields the collapse transition around $\epsilon_{nn} = 1~k_\text{B} T$ as shown in \Fig.~\ref{fig:fig2}(b), resulting in qualitative agreement of the theory with the simulation results.

\section{Results and Discussion}

\subsection{Cosolute-free networks}

\subsubsection{Volume transition}

\begin{figure}[t]
\centering
\includegraphics[width = 0.45\textwidth]{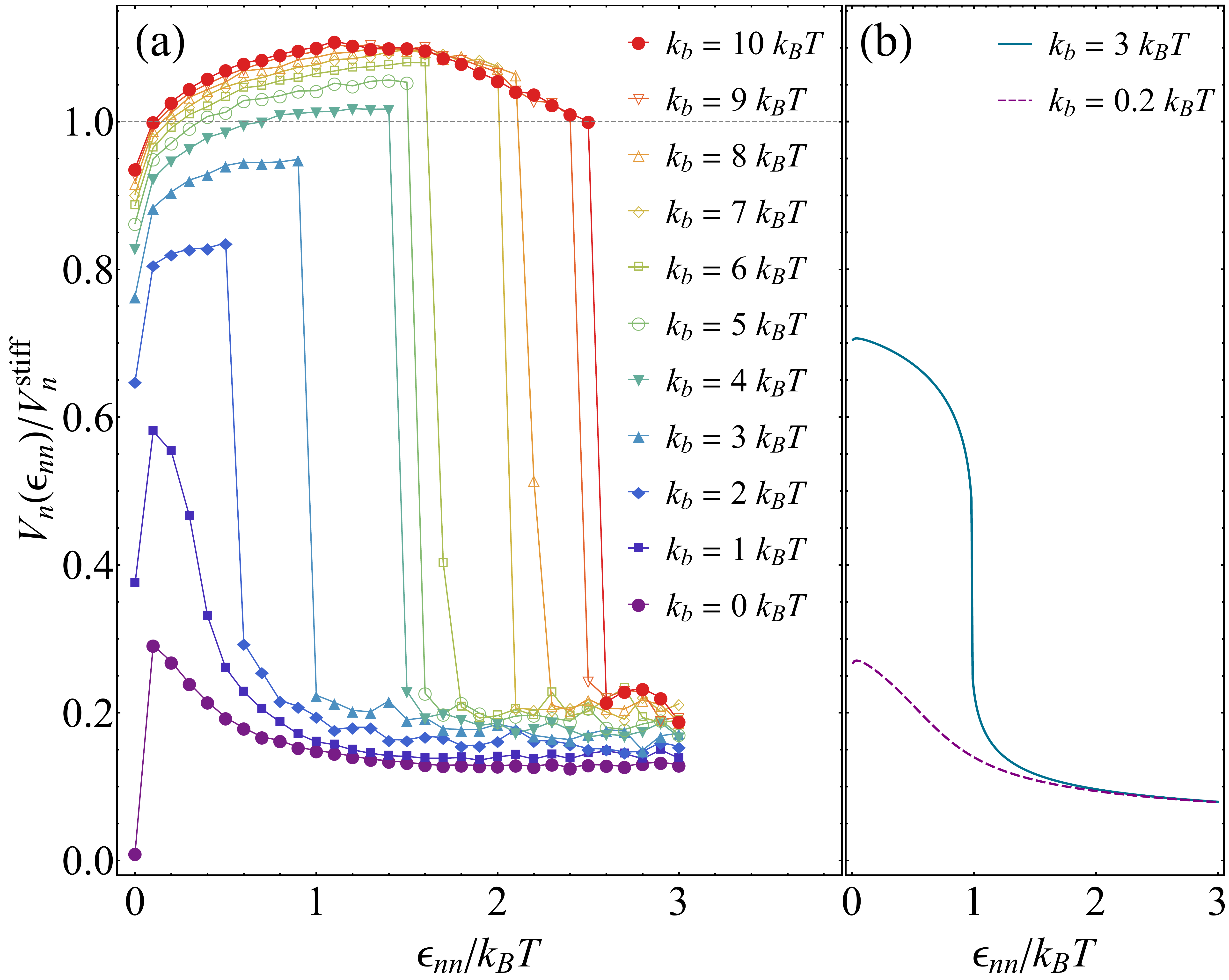}
\caption{Mean network volumes $V_{n}(\epsilon_{nn}) /V_{n}^\text{stiff}$ depending on solvent quality parameter $\epsilon_{nn}$ for various network bending rigidities $k_b$.
(a) Simulation results. (b) Theoretical predictions for semi-flexible ($k_b/k_\text{B} T = 3$) network and flexible ($k_b/k_\text{B} T = 0.2$) network, based on the mean-field theory in Section~~\ref{sec:meanfield}.
}\label{fig:fig2}
\end{figure}

	Before we present results on cosolute partitioning we briefly discuss the network properties for the cosolute-free reference systems. Figure~\ref{fig:fig2} shows the mean network volume $V_{n}(\epsilon_{nn}) \equiv \langle A(\epsilon_{nn}) d(\epsilon_{nn}) \rangle$, where $A(\epsilon_{nn})$ is the area in the $xy$ plane and $d(\epsilon_{nn})$ is the thickness of the film, respectively.  The mean volume is rescaled by $V_{n}^\text{stiff}$, the mean volume of a stiff and repulsive reference network with $k_b = 10~k_\text{B} T$ and $\epsilon_{nn} = 0.1~k_\text{B} T$. 	The mean volume exhibits dramatic changes with varying $k_b$ and $\epsilon_{nn}$. 
	For $\epsilon_{nn}=0$,  the equilibrium volumes increase with $k_b$, which can be readily understood by the larger stiffness of the network chains.
	As $\epsilon_{nn}$ increases from zero (i.e., from perfectly ideal polymer to a good solvent state), $V_{n}$ substantially increases overall due to the excluded repulsion between the neighboring network beads. The larger the network bending rigidities the smaller the effect.  As the effective interaction parameter $\epsilon_{nn}$ further increases (moving towards a bad solvent), there are two opposing effects that result in two different trends depending on the chain rigidity: First, the attraction depth of the LJ interaction increases, but secondly also the excluded volume repulsion slightly increases with growing $\epsilon_{nn}$ (since it is also a prefactor of the repulsive LJ term). For the less flexible chains ($k_b\gtrsim 2~k_\text{B} T$), only little inter-bead interactions are present, mostly only bonded, nearest-neighbor interactions. In this case, the excluded-volume effect dominates in our model and the network swells slightly. For the more flexible systems ($k_b\lesssim 2~k_\text{B} T$) there are more many-bead interactions and the overall increased attraction outweighs the steric exclusion, as expected for a simple LJ fluid. 


	More importantly, further increase of $\epsilon_{nn}$ leads to a discontinuous collapse transition for semi-flexible and stiff chains ($k_b\gtrsim 2~k_\text{B} T$).  This abrupt drop-down of the polymer volume \cite{tanaka1977critical, Schneider2004} is correlated to the breakdown of the network structure. The critical attraction for the collapse increases from $\epsilon_{nn} \approx 0.6$  to $2.6~k_\text{B} T$ for the stiffest chains. For very large $\epsilon_{nn}$, all networks saturate almost to the same (closed-packed) limit  around $V_{n} /V_{n}^\text{stiff} = 0.2$.  
For the semi-flexible network film, where the collapse transition occurs around the transition point $\epsilon_{nn} = 1 ~k_\text{B} T$ (\Fig~\ref{fig:fig2}), the volume fluctuations are extremal around the transition point (see Figure S7 in SI), as qualitatively expected from experimental knowledge~\cite{tanaka1977critical,Habicht2015}. The transition is continuous for the more flexible chains, as known from previous theories for polymer collapse transitions, either on a single chain level \cite{grosberg1992quantitative} or the network level using the Flory rubber elasticity \cite{khokhlov1980swelling,erman1986critical,khokhlov1993conformational} and combination of the bending energy and Flory--Huggins theory in the rod limit \cite{barenbrug1995highly}.  The transition behavior observed in the simulations can also be semi-quantitatively reproduced by the  mean-field theory (without cosolutes), as shown in \Fig~\ref{fig:fig2}(b).   In particular,  the theory clearly shows the effect of the gel semi-flexibility on the sharp volume transition, i.e., as in the simulations the transition is continuous for the flexible chains, and becomes discontinuous for semi-flexible chains. This is in fact also consistent with experiments of PNIPAM networks whose volume transitions are discontinuous above a certain cross-linking degree which can not be described by simple Gaussian mean-field theories (like Flory--Huggins) and was ascribed to both flexibility and defect effects~\cite{Tanaka1992}. Our results suggest that solely a diminished flexibility is already sufficient to push a strongly cross-linked system to discontinuous volume transitions.  
	
\subsubsection{Monomer packing fraction profiles}\label{sec:pack0}

\begin{figure}[h]
\centering
\includegraphics[width = 0.35\textwidth,trim={0 1.8cm 0 1.8cm},clip]{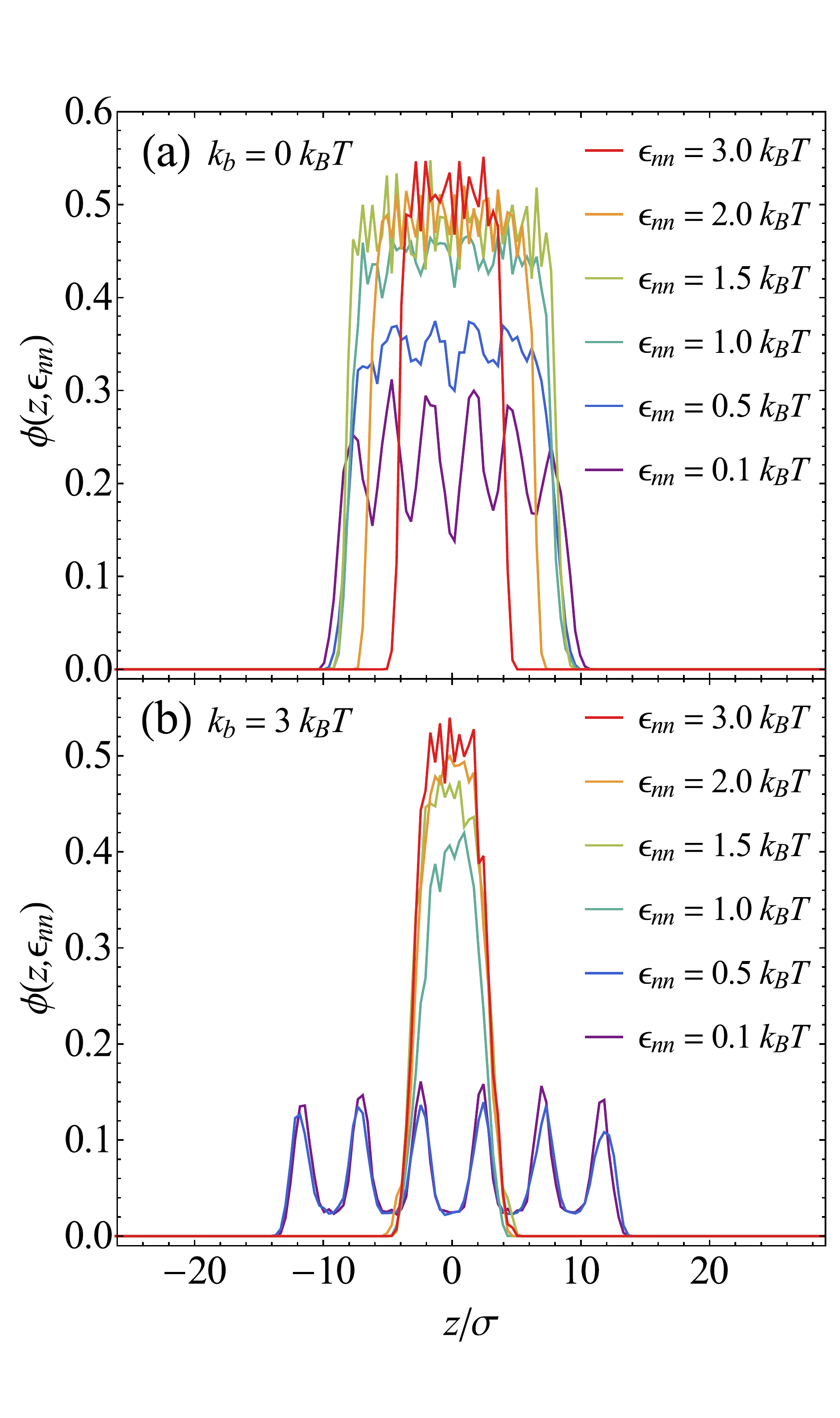}
\caption{Mean network monomer density profiles (packing fraction) $\phi(z, \epsilon_{nn})$ for (a) flexible ($k_b/k_\text{B} T = 0$) and (b) semi-flexible ($k_b / k_\text{B} T = 3$) polymers at various values of $\epsilon_{nn}$, see the legend.
}\label{fig:fig4}
\end{figure}

	\begin{figure*}[h!]
\centering
\includegraphics[width =0.9\textwidth]{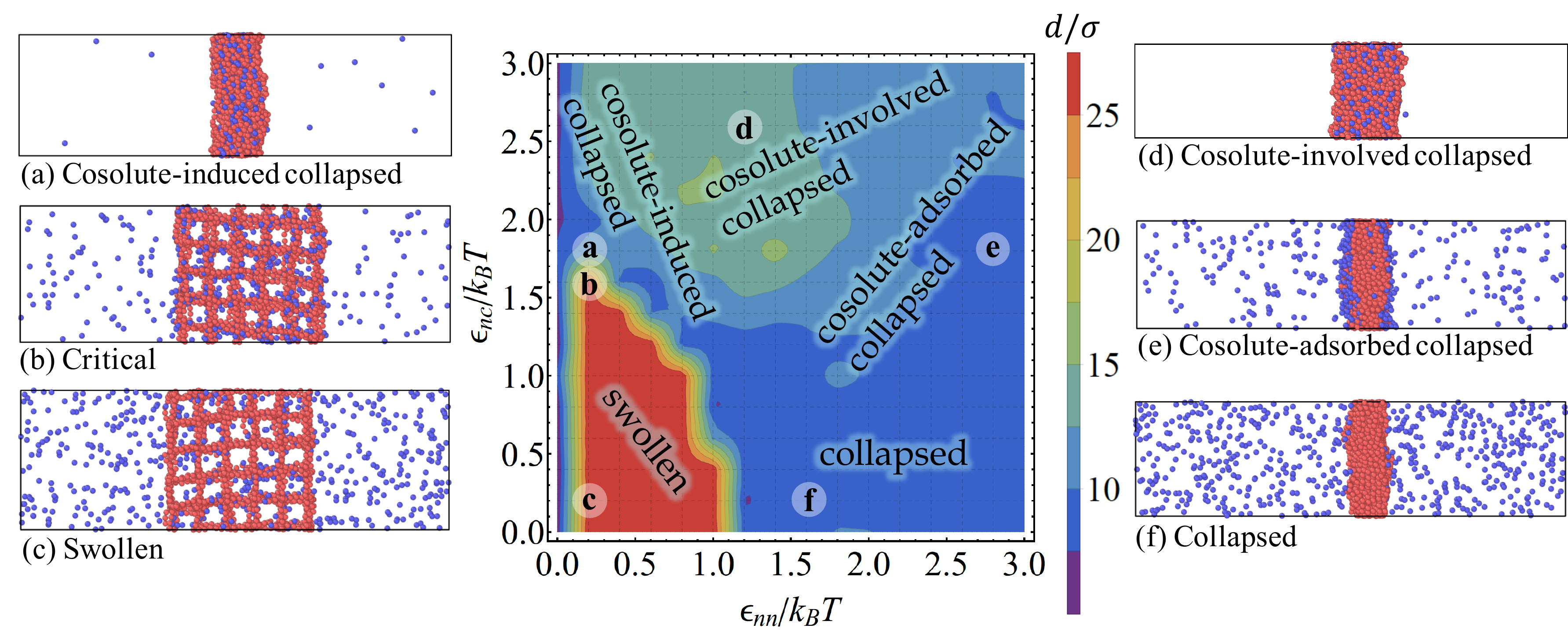}
\caption{
Center: The mean width $d$ of the semi-flexible network film, as a function of both the solvent quality parameter $\epsilon_{nn}$ and the cosolute coupling parameter $\epsilon_{nc}$. 
The regions on the landscape are classified into ``swollen'', ``collapsed'', ``cosolute-induced collapsed", ``cosolute-involved collapsed'', and ``cosolute-adsorbed collapsed'' states illustrated by representative simulation snapshots:
(a) The cosolute-induced collapsed state in a good solvent ($\epsilon_{nn} = 0.2 ~k_\text{B} T$) and strong cosolute coupling ($\epsilon_{nc} = 1.8 ~k_\text{B} T$) conditions.  The collapse transition is induced by high attraction between the network and the cosolute particles although the network intrinsically favors the swollen structure.
(b) The critical swollen state in a good solvent ($\epsilon_{nn} = 0.2 ~k_\text{B} T$) and less strong cosolute coupling ($\epsilon_{nc} = 1.6 ~k_\text{B} T$) conditions.
(c) The swollen state in a good solvent ($\epsilon_{nn} = 0.2 ~k_\text{B} T$) and weak cosolute coupling ($\epsilon_{nc} = 0.2 ~k_\text{B} T$).
(d) The cosolute-involved collapsed state in a bad solvent ($\epsilon_{nn} = 1.2 ~k_\text{B} T$) and very strong cosolute coupling ($\epsilon_{nc} = 2.6 ~k_\text{B} T$) conditions. 
Due to the large attraction between the cosolutes and the network, the film width is larger than the other collapsed states.
(e) The cosolute-adsorbed collapsed state in a very bad solvent ($\epsilon_{nn} = 2.8 ~k_\text{B} T$) and strong cosolute coupling ($\epsilon_{nc} = 1.8 ~k_\text{B} T$) conditions.
The collapse transition is induced by high attraction between network particles, yielding the cosolute particle adsorption onto the network surfaces.
(f) The collapsed structure in a bad solvent ($\epsilon_{nn} = 1.6 ~k_\text{B} T$) and weak cosolute coupling ($\epsilon_{nc} = 0.2 ~k_\text{B} T$).
}\label{fig:fig5}
\end{figure*}

	The collapse transition of the network is also clearly visible in the local packing fraction resolved in $z$, $\phi(z, \epsilon_{nn}) = v c_n (z,\epsilon_{nn})$,
where $v = \pi \sigma^3 /6$ represents the volume of a single bead, and $c_n (z,\epsilon_{nn})$  is the local number density.
 In \Fig~\ref{fig:fig4}, $\phi(z)$ for the semi-flexible and flexible network films is shown for various values of the solvent quality parameter $\epsilon_{nn}$, respectively. The packing fraction is overall transformed from a structured shape with salient multiple peaks into a predominant single broad peak, signifying the collapsed state  after $\epsilon_{nn}$ crosses the volume transition value. However, as discussed above, the flexible network (\Fig~\ref{fig:fig4}(a)) collapses rather continuously and maintains a characteristically ordered  structure even up to $\epsilon_{nn} = 1 ~k_\text{B} T$, whereas the semi-flexible network structure collapses abruptly (\Fig~\ref{fig:fig4}(b)), and a peak fluctuates around the center $z=0$.
Note that the integral over the profiles yields the monomer number density per area and is not a conserved quantity 
 as the aspect ratio of the network (i.e., the ratio of the length in $x$ or $y$ to the width in $z$) depends on the flexibility and $\epsilon_{nn}$. 
	
	\subsection{Networks including cosolutes}

We now consider effects of interacting cosolute particles. Hence, in addition to the intrinsic solvent quality parameter $\epsilon_{nn}$, we now study the influence  of varying the network--cosolute interactions given by $\epsilon_{nc}$, i.e.,  the cosolute coupling parameter.

\subsubsection{Network-cosolute state diagram}

	Figure~\ref{fig:fig5} shows in the central panel (colored landscape) the mean film width $d/\sigma$ of the semi-flexible network in dependence of both interaction parameters along with representative simulation snapshots. We now study the width instead of the total volume to distinguish better the effects on the aspect ratio of the network. 
	We see that there are overall five regions and a critical line that characterizes the collapse transition. We classify these regions into ``swollen'', ``collapsed'', ``cosolute-induced collapsed'',``cosolute-adsorbed collapsed", and ``cosolute-involved collapsed'' states, as depicted on the landscape. Hence, in addition to the intrinsic volume transition between the ``swollen'' (c) and the ``collapsed'' (f) states, on which the rapid color change from the red (swollen) region to the blue (collapsed) region is notable as yellow `critical' line (b), we observe three new cosolute-related states. 

		 Most importantly, we find cosolute-induced collapsed states (a), in which even in a good solvent condition, the network can undergo a sharp collapse transition triggered by the strong cosolute attraction to the monomers. The reason for this cosolute-induced collapse is a high energy gain of the system by bridging cosolutes located between multiple attracted monomers. The existence of those has been demonstrated before only for single chains by computer simulations and statistical theory~\cite{ Heyda2013,mukherji2013coil,mukherji2014polymer,rodriguez2015mechanism,rodriguez2015urea} as well as experiments.~\cite{rika1990intermolecular,lee1997effects,heyda2017guanidinium}  Evidently, this effect also plays a role in polymer networks. 

	
	In the ``cosolute-involved'' collapsed state (d), where the solvent quality ranges from $\epsilon_{nn} \approx 0.2~k_\text{B} T$ to $\epsilon_{nn} \approx 1.6~k_\text{B} T$ and the cosolute coupling is strong, $\epsilon_{nc} > 1.8~k_\text{B} T$, the network is intrinsically in the collapsed regime but also embeds the attracted cosolutes. 
In this state, the gel volume becomes approximately two times larger than the collapsed volume in a bad solvent without cosolutes embedded.
Finally, in the ``cosolute-adsorbed collapsed" state (e), where both network--network and network--cosolute attractions are strong, the network is collapsed and cosolutes are attracted to the surface of the network but do not penetrate it.  However, these two states ``cosolute-involved collapsed''  and ``cosolute-adsorbed collapsed" are not clearly distinguishable as it is not well defined  in which parameter region they are actually equilibrium states.  Their exact stability regions may depend on the simulation history and could originate from kinetically trapped trajectories due to the strong attractions involved.

\subsubsection{Impact of cosolutes on the volume transition}

\begin{figure*}
\centering
\includegraphics[width = 1\textwidth]{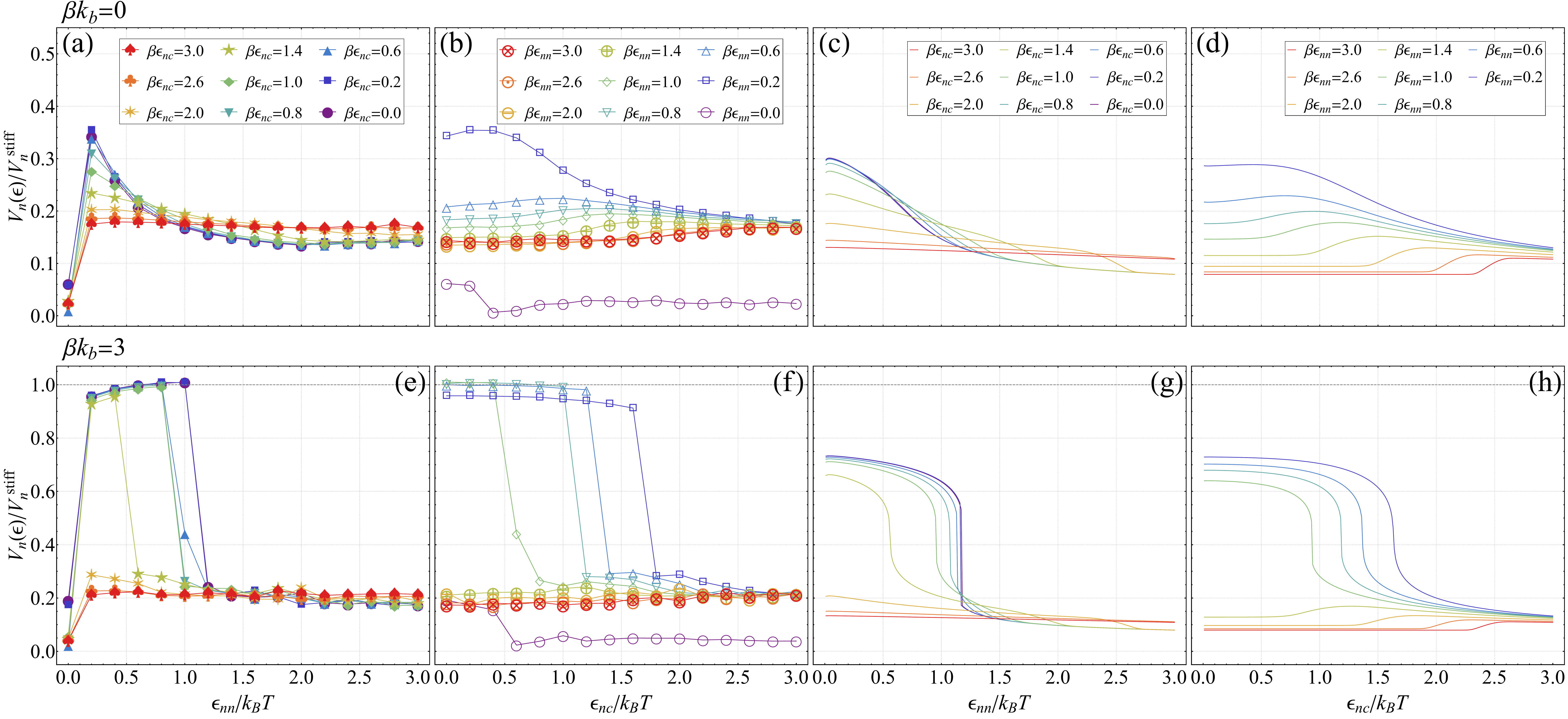}
\caption{Simulation results for mean network film volumes $V_{n}(\epsilon) /V_{n}^\text{stiff}$ depending on (a) $\epsilon_{nn} / k_\text{B} T$, (b) $\epsilon_{nc} / k_\text{B} T$ for the flexible network film, (e) $\epsilon_{nn} / k_\text{B} T$, and (f) $\epsilon_{nc} / k_\text{B} T$ for the semi-flexible network film.
Theoretical predictions as functions of (c) $\epsilon_{nn} / k_\text{B} T$, (d) $\epsilon_{nc} / k_\text{B} T$ for the flexible network film, (g) $\epsilon_{nn} / k_\text{B} T$, and (h) $\epsilon_{nc} / k_\text{B} T$ for the semi-flexible network film.
The mean-field theory (\Eqs~\eqref{eq:free_energy}--\eqref{eq:eq12}), as a function of the solvent quality parameter $\epsilon_{nn}$ for flexible networks ($l_p=0.43\sigma$, upper panels) and semi-flexible networks ($l_p=6\sigma$, bottom panels) is used.
}\label{fig:fig6}
\end{figure*}

	In \Fig~\ref{fig:fig6} we plot the mean network volume $V_n(\epsilon_{nn},\epsilon_{nc})$ in a more quantitive representation for both flexible and semi-flexible models versus $\epsilon_{nn}$ (for various $\epsilon_{nc}$) or versus $\epsilon_{nc}$ (for various $\epsilon_{nn}$). 
	For the flexible network, disregarding the jump when going from a non-interacting to interacting polymer at small $\epsilon_{nn}\simeq 0.1~k_\text{B} T$,  the volume decreases continuously without any sharp transition irrespective of the cosolute interaction, see panel (a). For fixed solvent quality, only in very good solvents, $\epsilon_{nn}\simeq 0.2~k_\text{B} T$, the change with cosolute interaction is substantial, see panel (b), where large network--cosolute attractions strongly collapse the network. For poorer solvents the network is always collapsed without much change induced by the cosolutes.  For the semi-flexible network film, panels (e) and (f), the sharp volume transitions are recovered. The transition shifts to smaller network attractions $\epsilon_{nn}$ for larger cosolute attraction, see panel (e). This is a signature of the cosolute-induced collapse we discussed above. The same signature can be found for good solvents in panel (f). We note that we also observe that the volume fluctuations of semi-flexible network films are maximized around the respective transition lines and that cosolute fluctuations inside the network are intimately coupled to them, see SI.

The features and trends reported in the simulations are also captured by the mean-field theory presented in \Eqs~\eqref{eq:free_energy}--\eqref{eq:f_bulk}. \Figures~\ref{fig:fig6}(c) and (d) show, respectively, the network volume predicted by this theory as a function of $\epsilon_{nn}$ and $\epsilon_{nc}$ for flexible chains. We observe that, including non-interacting cosolutes in the system ($\epsilon_{nc}=0$) leads to the swelling of the network 
due to the ideal gas contribution to the pressure induced by the density of the non-interacting cosolutes, which has to be compensated by the expanded network to keep the overall pressure constant. This swelling is emphasized when we increase the cosolute--monomer repulsion ($\epsilon_{nc}\approx 0.2~k_\text{B} T$). A similar finding has been reported for the swelling of charged microgels in salty media when the excluded-volume repulsive interactions exerted by the polymer network are taken into consideration.~\cite{moncho2016swelling} However, as soon as we increase $\epsilon_{nc}$ beyond $0.6~k_\text{B} T$, the monomer--cosolute attraction leads to a progressive shrinking of the network for small values of $\epsilon_{nn}$. This is a clear evidence of the cosolute-induced collapse, by which a network in good solvent conditions (swollen) tends to collapse in the presence of very attractive cosolutes. 

On the other hand, if we consider an already collapsed network in a bad solvent (moderate and large values of $\epsilon_{nn}$), for relatively small values of $\epsilon_{nc}$ the cosolutes tend to be excluded from the network, and so they are expected to surface-adsorb (although the theoretical model is not able to study the adsorption, it does predict the exclusion from the network). However, further enhancement of network-cosolute attraction leads finally to an increase of the network volume. In this case, the collapsed network is forced to swell to permit the strongly attractive cosolutes to diffuse inside.

\Figure~\ref{fig:fig6}(d) provides a very instructive illustration of the role of the cosolute network interaction. Increasing $\epsilon_{nc}$ leads first to an increase of the network volume. Indeed, for moderate attractions, the network swells to increase the contact surface exposed by the polymer chains allowing the attractive cosolutes to diffuse inside. However, if we increase more $\epsilon_{nc}$, the multiple contacts between monomers and cosolutes finally induce the network collapse with the cosolutes embedded inside; the so-called cosolute-involved collapse. This leads to the appearance of a maximum that, although is only weakly established in the simulations, it is in fact a known effect~\cite{Heyda2013}.

\Figures~\ref{fig:fig6}(g) and (h) depict again the swelling response of the polymer network but for semi-flexible chains. As observed, the simulation and theory swelling and transition trends with cosolutes are fully consistent. Indeed, the theory predicts a shift of the sharp volume transition to larger values of $\epsilon_{nn}$ for effective cosolute-monomer repulsions. Then, by increasing $\epsilon_{nc}$ above $0.3~k_\text{B} T$, the transition moves to smaller values of $\epsilon_{nn}$, indicating that the attractive cosolute is causing the cosolute-induced collapse. For $\epsilon_{nc} \gtrsim 1.5~k_\text{B} T$ the attraction is so strong that it is able to fully collapse the network even for good solvent conditions. Moreover, for moderate to large values of $\epsilon_{nn}$ the theory also predicts the cosolute-involved collapse reported in the simulations for intense cosolute--monomer attractions.


\subsubsection{Cosolute packing fraction profiles}

\begin{figure*}
\centering
\includegraphics[width = 0.85\textwidth]{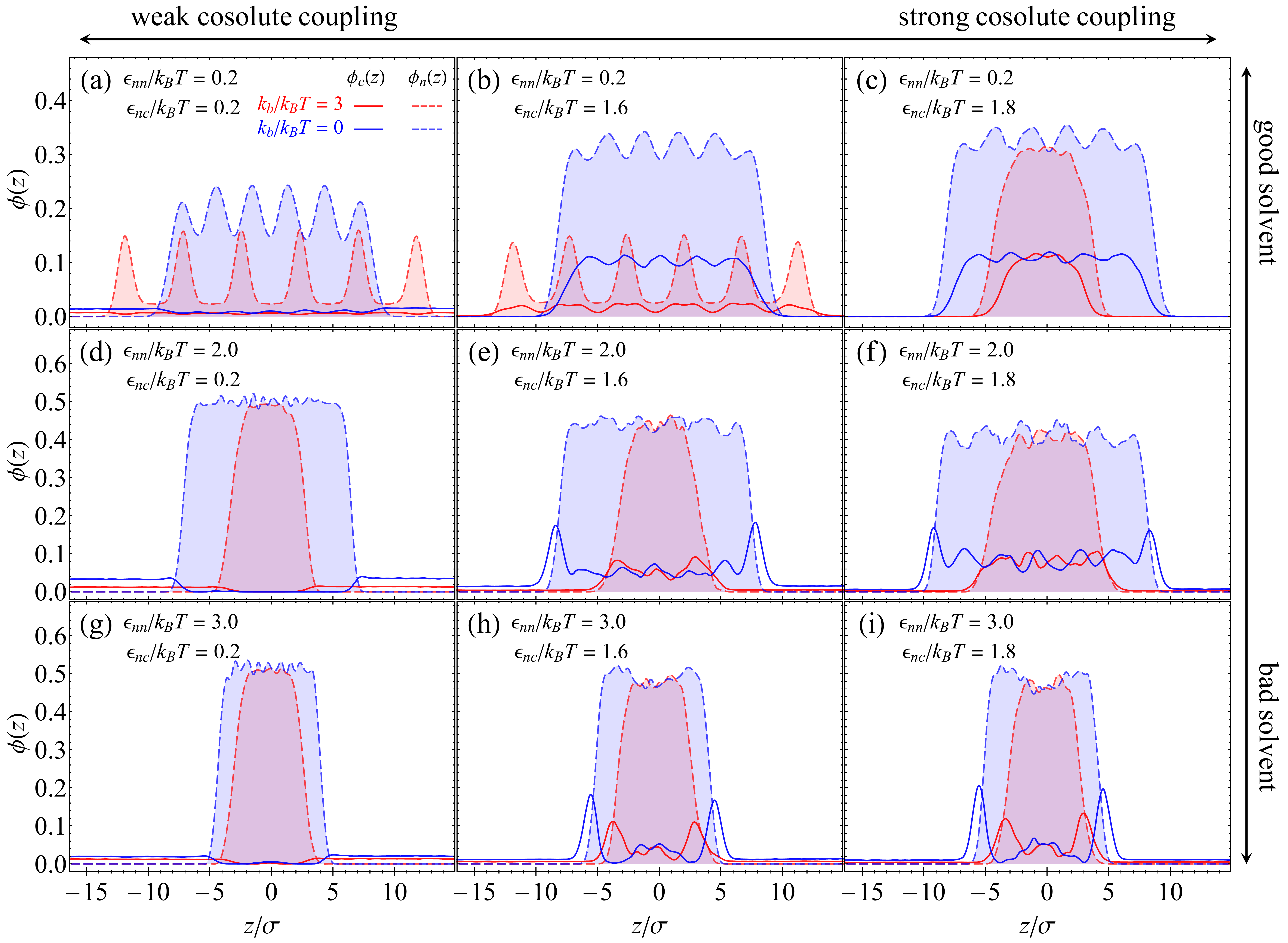}
\caption{Mean local packing fractions $\phi_{n}(z)$ of network particles (dashed lines) and $\phi_{c}(z)$ of cosolute particles (solid lines), for semi-flexible (red) and flexible (blue) network films, at various values of the interaction parameters $\epsilon_{nn}$ and $\epsilon_{nc}$. 
}\label{fig:fig7}
\end{figure*}

	The mean local packing fractions $\phi_{n}(z)$ of the network particles and $\phi_{c}(z)$ of the cosolutes are shown in \Fig~\ref{fig:fig7}, for the semi-flexible (red) and flexible (blue) network films. In this matrix representation, the network solvent quality decreases from top to bottom and the cosolute coupling increases from left to right. Hence, 
	at the bottom we see signatures of more collapsed states, and moving to the right we see more cosolutes at or inside the network. Some details, however, are rather complex. For instance, a cosolute-induced collapse transition is found  in (b)$\rightarrow$(c) for the semi-flexible films, in contrast to the cosolute-induced transition for the flexible films: The transition (d)$\rightarrow$(f) is rather a swelling transition where the cosolutes inside the flexible network strengthen the network structure. The latter effect is due to a collective adsorption of the cosolutes accumulating particularly at the network cross-linkers \cite{Johan2016}, which in conjugation with the gel flexibility maintains the network structure.   In the weak cosolute coupling regime, as shown in \Figs~\ref{fig:fig7}(a), (d) and (g), the cosolutes favor predominantly being outside the network film, throughout the entire collapse transition due to the change of the solvent quality. A combination of the bad solvent quality and the strong cosolute coupling, as shown in \Figs~\ref{fig:fig7}(e), (f), (h) and (i), enhances the cosolutes cumulation not only into the network film but also on the interface between the film and the bulk, leading to a high surface adsorption effect.
	
\subsubsection{Cosolute global partitioning}

We finally  investigate the cosolutes' global partitioning. It is computed using the definition $\mathcal{K}(\epsilon_{nn},\epsilon_{nc}) = \phi_{c}^\text{in} / \phi_{c}^\text{out}$
where the mean cosolute packing fraction inside the film is computed as $\phi_{c}^\text{in} = \int_{\rm in} \text{d}z \phi_{c}(z) / \Delta l_\text{in}$ integrating over the width $\Delta l_\text{in}$ of the film, and the mean cosolute packing fraction outside the network is computed as $\phi_{c}^\text{out} = \int_{\rm out} \text{d}z \phi_{c}(z) /(L_z - \Delta l_\text{in} - 2 l_\text{surf})$ over the range outside the network.
	Here we subtract the surface effects on the cosolutes by carefully choosing the surface width $l_\text{surf}$ around the film--bulk interfaces, respectively depending on $\epsilon_{nn}$ and $\epsilon_{nc}$ (see \Figs~S4 and S5 in SI).  
	We note that we also computed the fluctuations of the local packing fractions $\Delta \phi_{n,c}^{2}(z)$, i.e., the variance of $\phi_{n,c}(z)$, to obtain partitioning fluctuations. (see \Fig~S6 in SI).

\begin{figure}
\centering
\includegraphics[width = 0.5\textwidth]{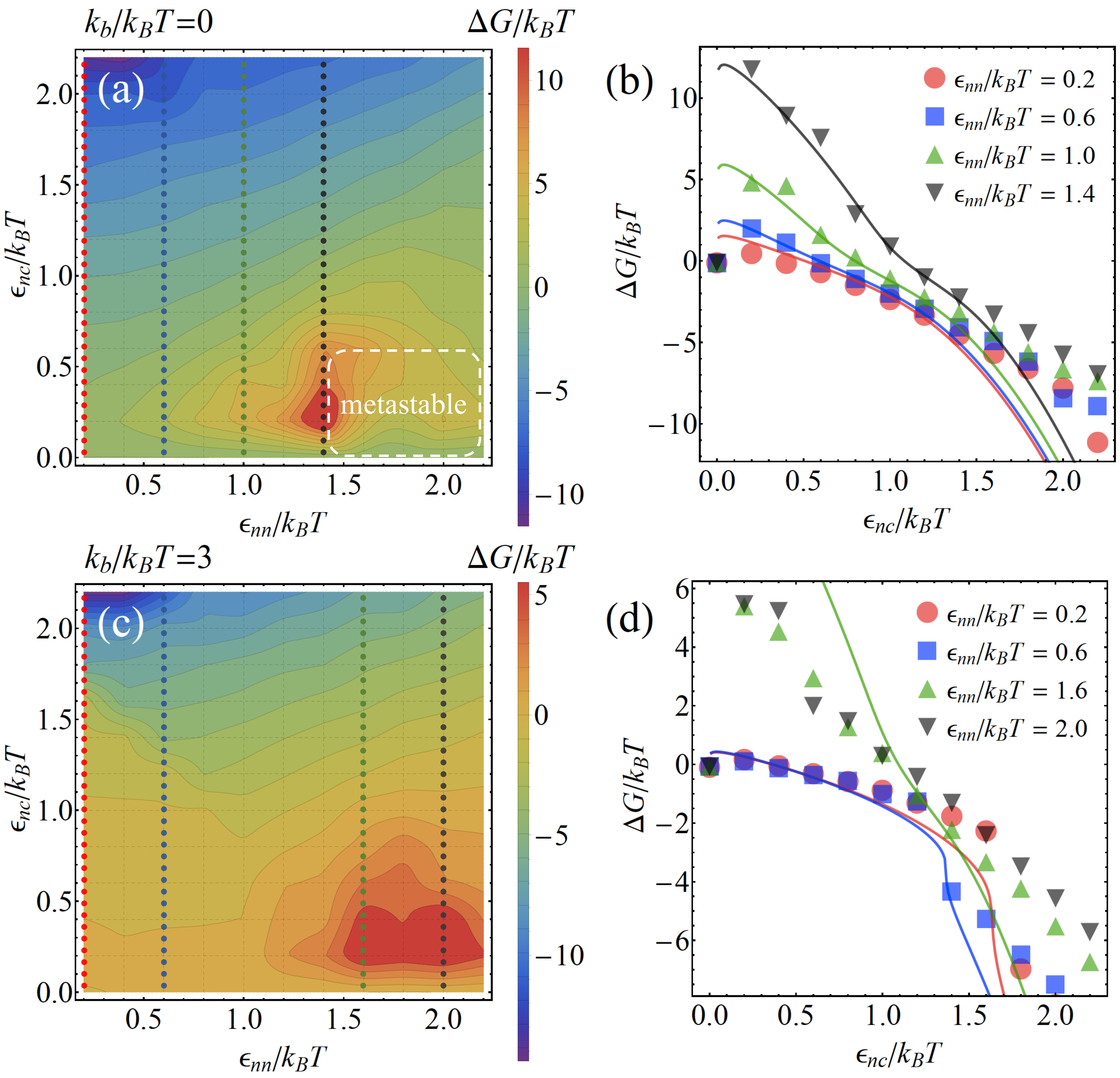}
\caption{
For flexible network films: 
(a) Adsorption energy profile $\Delta G (\epsilon_{nn}, \epsilon_{nc})$, 
(b) $\Delta G (\epsilon_{nc})$ (symbols) for different values of $\epsilon_{nn} = 0.2, 0.6, 1.0,$ and $1.4 ~k_\text{B} T$. The solid lines depict the theory prediction for $\Delta G$.
For semi-flexible network films: 
(c) Adsorption energy profile $\Delta G (\epsilon_{nn}, \epsilon_{nc})$, 
(d) $\Delta G (\epsilon_{nc})$ (symbols) for different values of $\epsilon_{nn} = 0.2, 0.6, 1.6,$ and $2 ~k_\text{B} T$. The solid lines depict the theory prediction for $\Delta G$.
The dotted lines in (a) and (c) denote respective fixed values of $\epsilon_{nn}$ used in (b) and (d).
}\label{fig:fig8}
\end{figure}

The partitioning $\mathcal{K}$ is related to the transfer free energy $\Delta G = - k_\text{B} T \ln \mathcal{K}$, a measure of the effective cosolute--network interaction.
We show $\Delta G (\epsilon_{nn},\epsilon_{nc})$ in \Figs~\ref{fig:fig8}(a) and (c). The partitioning depends not only on the cosolute coupling but also significantly on the solvent quality. The cosolute uptake is largely affected by the solvent quality, particularly when the film collapses in a bad solvent and with weak cosolute couplings.
For the two different network film models, $\Delta G$ shows different shapes of the landscapes, reflecting significant role of the network flexibility.
The transfer energy overall becomes attractive as the cosolute coupling increases, however, $\Delta G$ for the semi-flexible film has a large zero-energy region, particularly in the swollen zone (see \Fig~\ref{fig:fig5}), which is not seen in $\Delta G$ for the flexible films.
The slow relaxation of the trapped cosolutes in the flexible network films at $\epsilon_{nn} > 1.4 ~k_\text{B} T$ in the weak cosolute coupling results in a strong cosolute trapping effect on $\Delta G$ with maximal repulsion.
In this case, the flexible network collapses exceedingly faster than the cosolutes relaxation, thus the cosolutes are trapped in the collapsed network for a long time (beyond our total simulation time) before they are equilibrated by diffusing out of the network, i.e., crossing over the large barrier to an energy minimum.
We depict this metastable parameter zone which gives rise to the strong cosolute trapping effect by a dashed line in \Fig~\ref{fig:fig8}(a).
Note that the gel collapse transition can occur at both extrema of $\Delta G$.
 

Figures~\ref{fig:fig8}(b) and (d) show the transfer free energy $\Delta G$ depending on the cosolute coupling $\epsilon_{nc}$ for different values of the solvent quality parameter $\epsilon_{nn}$, depicted by dotted lines in panels (a) and (c). As the cosolute coupling increases, $\Delta G$ decreases, particularly signalling the cosolute-induced collapse in a good solvent condition. For the semi-flexible film this reduction of $\Delta G$ shown in \Fig~\ref{fig:fig8}(d) is dramatic, that is, abruptly decreasing around the collapse transition point in a good solvent condition.  
 
The simple mean-field partitioning theory presented in Section~\ref{sec:meanfield} can also be applied to determine the ratio of concentrations of cosolutes inside and outside the network and, from them, the desired transfer free energy  $\beta \Delta G =-\ln(c_{c}^\text{in} / c_{c}^\text{out})$. The results for $\Delta G (\epsilon_{nc})$ are shown by solid lines in \Figs~\ref{fig:fig8}(b) and (d) for flexible and semi-flexible networks, respectively. Good agreement of the theoretical prediction with the simulation results is found. In general, the theoretical results confirm that increasing the network--cosolute attraction leads to an increase of the cosolute concentration, which corresponds to more negative adsorption energies for both, flexible and stiff networks. For the case of a polymer network in a bad solvent, the gel phase is already collapsed. In this particular situation, cosolutes are strongly excluded from the gel phase as they disrupt the attractive bonds between neighboring monomers. As a result of this, very large and positive transfer free energies are predicted in this regime, and only for relative large cosolute--network attractions (about $\epsilon_{nc} \gtrsim 1~k_\text{B} T$) the transger free energy becomes negative. This effect is also well captured by the mean-field theory, in accordance with the simulation data (see solid lines in \Figs~\ref{fig:fig8}(b) and (d)). Moreover, in some cases the theory shows to be remarkably good, as it even provides quantitative agreement with the simulation results, especially for moderate inter-monomer attractions (see for instance that the model is able to capture the abrupt decrease of the free energy induced bt the cosolute attraction, achieved for semi-flexible networks in good solvent conditions).

\section{Summary and concluding remarks}\label{sec:conclusion}

In summary, we studied the partitioning of model cosolutes to a simple cubic polymer network film for a wide range of solvent qualities and interactions by using extensive coarse-grained computer simulations, with special focus on the influence of polymer flexibility and volume transitions. In addition, the computer simulation results were corroborated by the theoretical predictions for the network swelling and the cosolute partitioning obtained with a two-component mean-field approach that takes into account the chain flexibility, the excluded-volume effects and the role of the attractive interaction up to second order in the virial expansion. A sharp discontinuous collapse transition was found only in semi-flexible networks, in agreement with a newly developed mean-field theory.
Our simulations and theory are generic in a sense that the interaction parameters $\epsilon_{nn}$ and $\epsilon_{nc}$ consider multiple combinations of effective interactions, going from repulsive to attractive pair interactions.
 
In the presence of cosolutes we then found a rich and nontrivial topology of equilibrium gel structural states and partitioning despite the simplicity of the models. In particular, the network film width revealed that there are overall five distinct states that characterize a measure of the collapse transition, and we categorized these states into ``swollen'', ``collapsed'', and ``cosolute-induced'', ``cosolute-involved'', ``cosolute-adsorbed'' collapsed states.  The ``cosolute-induced'' collapsed state is noteworthy, showing a collective manner of the network collapse transition which is induced by the cosolutes in the strong cosolute coupling, even in good solvents, and accompanied by maximal but finite collective fluctuations.  In addition, the ``cosolute-involved'' collapsed state occurs in intermediate solvent conditions, and the ``cosolute-adsorbed'' collapsed state occurs in bad solvent conditions.

The global adsorption energy landscape computed via the cosolutes partitioning exhibits very different topologies on two different gel models, particularly showing the distinct swollen states in the semi-flexible gels.
For the flexible gel system the collapse transition driven by bad solvents and weak cosolute couplings results in a very strong trapping of the cosolutes inside the gel, giving rise to metastable states. Finally, we compared the volume transition and transfer free energy with a theoretical model. 
By using the mean-field level virial expansion, we found a good qualitative agreement of the theory with the simulation results.

The coarse-grained model based on the simple cubic network \cite{Aydt2000,Lu2002,pnetz1997,Erbas2015,Erbas2016,Li2016} entails further extension towards more realistic gel structures, e.g., multi-functional gels with random cross-linkings and defects.
Quantifying the partitioning and fluctuations in those systems depending on structural properties such as polydispersity \cite{Edgecombe2007}, as well as their coupling to cosolute diffusion and permeability\cite{gehrke}, would be a fecund future work. With these extensions of simulations as well as predictive theory the future rational design and development of soft functional materials will be hopefully facilitated. 

\begin{acknowledgements}
The authors thank Matthias Ballauff and Richard Chudoba for fruitful discussions.
This project has received funding from the European Research Council (ERC) under the European Union's Horizon 2020 research and innovation programme (grant agreement No.~646659-NANOREACTOR).
A.M.-J. acknowledge funding by the Spanish \textquoteleft Ministerio de Econom\'{\i}a y Competitividad (MINECO), Plan Nacional de Investigaci\'{o}n, Desarrollo e Innovaci\'{o}n Tecnol\'{o}gica (I+D+i)' (Project FIS2016-80087-C2-1-P).
The simulations were performed with resources provided by the North-German Supercomputing Alliance (HLRN).
\end{acknowledgements}

\setlength{\bibsep}{0pt}

\begin{mcitethebibliography}{91}
\providecommand*\natexlab[1]{#1}
\providecommand*\mciteSetBstSublistMode[1]{}
\providecommand*\mciteSetBstMaxWidthForm[2]{}
\providecommand*\mciteBstWouldAddEndPuncttrue
  {\def\EndOfBibitem{\unskip.}}
\providecommand*\mciteBstWouldAddEndPunctfalse
  {\let\EndOfBibitem\relax}
\providecommand*\mciteSetBstMidEndSepPunct[3]{}
\providecommand*\mciteSetBstSublistLabelBeginEnd[3]{}
\providecommand*\EndOfBibitem{}
\mciteSetBstSublistMode{f}
\mciteSetBstMaxWidthForm{subitem}{(\alph{mcitesubitemcount})}
\mciteSetBstSublistLabelBeginEnd
  {\mcitemaxwidthsubitemform\space}
  {\relax}
  {\relax}

\bibitem[Peppas \latin{et~al.}(2000)Peppas, Bures, Leobandung, and
  Ichikawa]{Peppas}
Peppas,~N.; Bures,~P.; Leobandung,~W.; Ichikawa,~H. Hydrogels in pharmaceutical
  formulations. \emph{Eur. J. Pharm. Biopharm.} \textbf{2000}, \emph{50},
  27--46\relax
\mciteBstWouldAddEndPuncttrue
\mciteSetBstMidEndSepPunct{\mcitedefaultmidpunct}
{\mcitedefaultendpunct}{\mcitedefaultseppunct}\relax
\EndOfBibitem
\bibitem[Hamidia \latin{et~al.}(2008)Hamidia, Azadia, and Rafiei]{Hamidi}
Hamidi,~M.; Azadi,~A.; Rafiei,~P. Hydrogel nanoparticles in drug delivery.
  \emph{Adv. Drug Delivery Rev.} \textbf{2008}, \emph{60}, 1638--1649\relax
\mciteBstWouldAddEndPuncttrue
\mciteSetBstMidEndSepPunct{\mcitedefaultmidpunct}
{\mcitedefaultendpunct}{\mcitedefaultseppunct}\relax
\EndOfBibitem
\bibitem[Tokarev and Minko(2010)Tokarev, and Minko]{Tokarev}
Tokarev,~I.; Minko,~S. Stimuli-Responsive Porous Hydrogels at Interfaces for
  Molecular Filtration, Separation, Controlled Release, and Gating in Capsules
  and Membranes. \emph{Adv. Mater.} \textbf{2010}, \emph{22}, 3446--3462\relax
\mciteBstWouldAddEndPuncttrue
\mciteSetBstMidEndSepPunct{\mcitedefaultmidpunct}
{\mcitedefaultendpunct}{\mcitedefaultseppunct}\relax
\EndOfBibitem
\bibitem[Stuart \latin{et~al.}(2010)Stuart, Huck, Genzer, M{\"u}ller, Ober,
  Stamm, Sukhorukov, Szleifer, Tsukruk, Urban, Winnik, Zauscher, Luzinov, and
  Minko]{stuart2010emerging}
Stuart,~M. A.~C.; Huck,~W.~T.; Genzer,~J.; M{\"u}ller,~M.; Ober,~C.; Stamm,~M.;
  Sukhorukov,~G.~B.; Szleifer,~I.; Tsukruk,~V.~V.; Urban,~M.; Winnik,~F.;
  Zauscher,~S.; Luzinov,~I.; Minko,~S. Emerging applications of
  stimuli-responsive polymer materials. \emph{Nat. Mater.} \textbf{2010},
  \emph{9}, 101--113\relax
\mciteBstWouldAddEndPuncttrue
\mciteSetBstMidEndSepPunct{\mcitedefaultmidpunct}
{\mcitedefaultendpunct}{\mcitedefaultseppunct}\relax
\EndOfBibitem
\bibitem[Lu and Ballauff(2011)Lu, and Ballauff]{Lu2011}
Lu,~Y.; Ballauff,~M. Thermosensitive core--shell microgels: from colloidal
  model systems to nanoreactors. \emph{Prog. Polym. Sci.} \textbf{2011},
  \emph{36}, 767--792\relax
\mciteBstWouldAddEndPuncttrue
\mciteSetBstMidEndSepPunct{\mcitedefaultmidpunct}
{\mcitedefaultendpunct}{\mcitedefaultseppunct}\relax
\EndOfBibitem
\bibitem[Welsch \latin{et~al.}(2010)Welsch, Ballauff, and
  Lu]{welsch2010microgels}
Welsch,~N.; Ballauff,~M.; Lu,~Y. \emph{Chemical design of responsive
  microgels}; Springer, 2010; pp 129--163\relax
\mciteBstWouldAddEndPuncttrue
\mciteSetBstMidEndSepPunct{\mcitedefaultmidpunct}
{\mcitedefaultendpunct}{\mcitedefaultseppunct}\relax
\EndOfBibitem
\bibitem[Lu \latin{et~al.}(2006)Lu, Mei, Drechsler, and
  Ballauff]{lu2006thermosensitive}
Lu,~Y.; Mei,~Y.; Drechsler,~M.; Ballauff,~M. Thermosensitive core--shell
  particles as carriers for Ag nanoparticles: modulating the catalytic activity
  by a phase transition in networks. \emph{Angew. Chem. Int. Ed.}
  \textbf{2006}, \emph{45}, 813--816\relax
\mciteBstWouldAddEndPuncttrue
\mciteSetBstMidEndSepPunct{\mcitedefaultmidpunct}
{\mcitedefaultendpunct}{\mcitedefaultseppunct}\relax
\EndOfBibitem
\bibitem[Ballauff and Lu(2007)Ballauff, and Lu]{lu2007thermosensitive}
Ballauff,~M.; Lu,~Y. ``Smart'' nanoparticles: preparation, characterization and
  applications. \emph{Polymer} \textbf{2007}, \emph{48}, 1815--1823\relax
\mciteBstWouldAddEndPuncttrue
\mciteSetBstMidEndSepPunct{\mcitedefaultmidpunct}
{\mcitedefaultendpunct}{\mcitedefaultseppunct}\relax
\EndOfBibitem
\bibitem[Menne \latin{et~al.}(2014)Menne, Pitsch, Wong, Pich, and
  Wessling]{Menne2014}
Menne,~D.; Pitsch,~F.; Wong,~J.~E.; Pich,~A.; Wessling,~M.
  Temperature-Modulated Water Filtration Using Microgel-Functionalized
  Hollow-Fiber Membranes. \emph{Angew. Chem. Int. Ed.} \textbf{2014},
  \emph{53}, 5706--5710\relax
\mciteBstWouldAddEndPuncttrue
\mciteSetBstMidEndSepPunct{\mcitedefaultmidpunct}
{\mcitedefaultendpunct}{\mcitedefaultseppunct}\relax
\EndOfBibitem
\bibitem[Plamper and Richtering(2017)Plamper, and Richtering]{Plamper2017}
Plamper,~F.~A.; Richtering,~W. Functional Microgels and Microgel Systems.
  \emph{Acc. Chem. Res.} \textbf{2017}, \emph{50}, 131--140\relax
\mciteBstWouldAddEndPuncttrue
\mciteSetBstMidEndSepPunct{\mcitedefaultmidpunct}
{\mcitedefaultendpunct}{\mcitedefaultseppunct}\relax
\EndOfBibitem
\bibitem[Wu and Yan(1994)Wu, and Yan]{wu1994studies}
Wu,~C.; Yan,~C.-Y. Studies of the swelling and drying kinetics of thin gelatin
  gel films by in situ interferometry. \emph{Macromolecules} \textbf{1994},
  \emph{27}, 4516--4520\relax
\mciteBstWouldAddEndPuncttrue
\mciteSetBstMidEndSepPunct{\mcitedefaultmidpunct}
{\mcitedefaultendpunct}{\mcitedefaultseppunct}\relax
\EndOfBibitem
\bibitem[Zhou and Wu(1996)Zhou, and Wu]{zhou1996situ}
Zhou,~S.; Wu,~C. In-situ interferometry studies of the drying and swelling
  kinetics of an ultrathin poly (N-isopropylacrylamide) gel film below and
  above its volume phase transition temperature. \emph{Macromolecules}
  \textbf{1996}, \emph{29}, 4998--5001\relax
\mciteBstWouldAddEndPuncttrue
\mciteSetBstMidEndSepPunct{\mcitedefaultmidpunct}
{\mcitedefaultendpunct}{\mcitedefaultseppunct}\relax
\EndOfBibitem
\bibitem[Lobaskin \latin{et~al.}(2012)Lobaskin, Bogdanov, and
  Vinogradova]{lobaskin2012interactions}
Lobaskin,~V.; Bogdanov,~A.~N.; Vinogradova,~O.~I. Interactions of neutral
  semipermeable shells in asymmetric electrolyte solutions. \emph{Soft Matter}
  \textbf{2012}, \emph{8}, 9428--9435\relax
\mciteBstWouldAddEndPuncttrue
\mciteSetBstMidEndSepPunct{\mcitedefaultmidpunct}
{\mcitedefaultendpunct}{\mcitedefaultseppunct}\relax
\EndOfBibitem
\bibitem[Petrache \latin{et~al.}(2006)Petrache, Tristram-Nagle, Harries,
  Ku{\v{c}}erka, Nagle, and Parsegian]{petrache2006swelling}
Petrache,~H.~I.; Tristram-Nagle,~S.; Harries,~D.; Ku{\v{c}}erka,~N.;
  Nagle,~J.~F.; Parsegian,~V.~A. Swelling of phospholipids by monovalent salt.
  \emph{J. Lipid Res.} \textbf{2006}, \emph{47}, 302--309\relax
\mciteBstWouldAddEndPuncttrue
\mciteSetBstMidEndSepPunct{\mcitedefaultmidpunct}
{\mcitedefaultendpunct}{\mcitedefaultseppunct}\relax
\EndOfBibitem
\bibitem[Leo \latin{et~al.}(1971)Leo, Hansch, and Elkins]{Leo1971}
Leo,~A.; Hansch,~C.; Elkins,~D. Partition coefficients and their uses.
  \emph{Chem. Rev.} \textbf{1971}, \emph{71}, 525--616\relax
\mciteBstWouldAddEndPuncttrue
\mciteSetBstMidEndSepPunct{\mcitedefaultmidpunct}
{\mcitedefaultendpunct}{\mcitedefaultseppunct}\relax
\EndOfBibitem
\bibitem[Gehrke \latin{et~al.}(1997)Gehrke, Fisher, Palasis, and Lund]{gehrke}
Gehrke,~S.; Fisher,~J.; Palasis,~M.; Lund,~M.~E. Factors determining hydrogel
  permeability. \emph{Ann. N. Y. Acad. Sci.} \textbf{1997}, \emph{831},
  179--207\relax
\mciteBstWouldAddEndPuncttrue
\mciteSetBstMidEndSepPunct{\mcitedefaultmidpunct}
{\mcitedefaultendpunct}{\mcitedefaultseppunct}\relax
\EndOfBibitem
\bibitem[Angioletti-Uberti \latin{et~al.}(2015)Angioletti-Uberti, Lu, Ballauff,
  and Dzubiella]{Stefano2015}
Angioletti-Uberti,~S.; Lu,~Y.; Ballauff,~M.; Dzubiella,~J. Theory of
  Solvation-Controlled Reactions in Stimuli-Responsive Nanoreactors. \emph{J.
  Phys. Chem. C} \textbf{2015}, \emph{119}, 15723--15730\relax
\mciteBstWouldAddEndPuncttrue
\mciteSetBstMidEndSepPunct{\mcitedefaultmidpunct}
{\mcitedefaultendpunct}{\mcitedefaultseppunct}\relax
\EndOfBibitem
\bibitem[Roa \latin{et~al.}(2017)Roa, Kim, Kandu\v{c}, Dzubiella, and
  Angioletti-Uberti]{Rafa2017}
Roa,~R.; Kim,~W.~K.; Kandu\v{c},~M.; Dzubiella,~J.; Angioletti-Uberti,~S.
  Catalyzed Bimolecular Reactions in Responsive Nanoreactors. \emph{ACS Catal.}
  \textbf{2017}, DOI: 10.1021/acscatal.7b01701.\relax
\mciteBstWouldAddEndPunctfalse
\mciteSetBstMidEndSepPunct{\mcitedefaultmidpunct}
{}{\mcitedefaultseppunct}\relax
\EndOfBibitem
\bibitem[Shibayama and Tanaka(1993)Shibayama, and Tanaka]{shibayama1993volume}
Shibayama,~M.; Tanaka,~T. \emph{Responsive gels: Volume transitions I};
  Springer, 1993; pp 1--62\relax
\mciteBstWouldAddEndPuncttrue
\mciteSetBstMidEndSepPunct{\mcitedefaultmidpunct}
{\mcitedefaultendpunct}{\mcitedefaultseppunct}\relax
\EndOfBibitem
\bibitem[Khokhlov(1980)]{khokhlov1980swelling}
Khokhlov,~A. Swelling and collapse of polymer networks. \emph{Polymer}
  \textbf{1980}, \emph{21}, 376--380\relax
\mciteBstWouldAddEndPuncttrue
\mciteSetBstMidEndSepPunct{\mcitedefaultmidpunct}
{\mcitedefaultendpunct}{\mcitedefaultseppunct}\relax
\EndOfBibitem
\bibitem[Erman and Flory(1986)Erman, and Flory]{erman1986critical}
Erman,~B.; Flory,~P. Critical phenomena and transitions in swollen polymer
  networks and in linear macromolecules. \emph{Macromolecules} \textbf{1986},
  \emph{19}, 2342--2353\relax
\mciteBstWouldAddEndPuncttrue
\mciteSetBstMidEndSepPunct{\mcitedefaultmidpunct}
{\mcitedefaultendpunct}{\mcitedefaultseppunct}\relax
\EndOfBibitem
\bibitem[Khokhlov \latin{et~al.}(1993)Khokhlov, Starodubtzev, and
  Vasilevskaya]{khokhlov1993conformational}
Khokhlov,~A.; Starodubtzev,~S.; Vasilevskaya,~V. \emph{Responsive gels: Volume
  transitions I}; Springer, 1993; pp 123--171\relax
\mciteBstWouldAddEndPuncttrue
\mciteSetBstMidEndSepPunct{\mcitedefaultmidpunct}
{\mcitedefaultendpunct}{\mcitedefaultseppunct}\relax
\EndOfBibitem
\bibitem[Barenbrug \latin{et~al.}(1995)Barenbrug, Smit, and
  Bedeaux]{barenbrug1995highly}
Barenbrug,~T.~M.; Smit,~J.; Bedeaux,~D. Highly swollen gels of semi-flexible
  polyelectrolyte chains near the rod limit. \emph{Polymer Gels and Networks}
  \textbf{1995}, \emph{3}, 331--373\relax
\mciteBstWouldAddEndPuncttrue
\mciteSetBstMidEndSepPunct{\mcitedefaultmidpunct}
{\mcitedefaultendpunct}{\mcitedefaultseppunct}\relax
\EndOfBibitem
\bibitem[Heskins and Guillet(1968)Heskins, and Guillet]{heskins1968solution}
Heskins,~M.; Guillet,~J.~E. Solution Properties of Poly(N-isopropylacrylamide).
  \emph{J. Macromol. Sci., Chem.} \textbf{1968}, \emph{2}, 1441--1455\relax
\mciteBstWouldAddEndPuncttrue
\mciteSetBstMidEndSepPunct{\mcitedefaultmidpunct}
{\mcitedefaultendpunct}{\mcitedefaultseppunct}\relax
\EndOfBibitem
\bibitem[Du{\v{s}}ek and Patterson(1968)Du{\v{s}}ek, and
  Patterson]{duvsek1968transition}
Du{\v{s}}ek,~K.; Patterson,~D. Transition in swollen polymer networks induced
  by intramolecular condensation. \emph{J. Polym. Sci. A-2 Polym. Phys}
  \textbf{1968}, \emph{6}, 1209--1216\relax
\mciteBstWouldAddEndPuncttrue
\mciteSetBstMidEndSepPunct{\mcitedefaultmidpunct}
{\mcitedefaultendpunct}{\mcitedefaultseppunct}\relax
\EndOfBibitem
\bibitem[Habicht \latin{et~al.}(2015)Habicht, Schmolke, Goerigk, Lange,
  Saalw{\"a}chter, Ballauff, and Seiffert]{Habicht2015}
Habicht,~A.; Schmolke,~W.; Goerigk,~G.; Lange,~F.; Saalw{\"a}chter,~K.;
  Ballauff,~M.; Seiffert,~S. Critical fluctuations and static inhomogeneities
  in polymer gel volume phase transitions. \emph{J. Polym. Sci. B Polym. Phys.}
  \textbf{2015}, \emph{53}, 1112--1122\relax
\mciteBstWouldAddEndPuncttrue
\mciteSetBstMidEndSepPunct{\mcitedefaultmidpunct}
{\mcitedefaultendpunct}{\mcitedefaultseppunct}\relax
\EndOfBibitem
\bibitem[Wu and Zhou(1997)Wu, and Zhou]{wu1997volume}
Wu,~C.; Zhou,~S. Volume phase transition of swollen gels: discontinuous or
  continuous? \emph{Macromolecules} \textbf{1997}, \emph{30}, 574--576\relax
\mciteBstWouldAddEndPuncttrue
\mciteSetBstMidEndSepPunct{\mcitedefaultmidpunct}
{\mcitedefaultendpunct}{\mcitedefaultseppunct}\relax
\EndOfBibitem
\bibitem[Schild \latin{et~al.}(1991)Schild, Muthukumar, and Tirrell]{conon}
Schild,~H.~G.; Muthukumar,~M.; Tirrell,~D.~A. Cononsolvency in mixed aqueous
  solutions of poly(N-isopropylacrylamide). \emph{Macromolecules}
  \textbf{1991}, \emph{24}, 948--952\relax
\mciteBstWouldAddEndPuncttrue
\mciteSetBstMidEndSepPunct{\mcitedefaultmidpunct}
{\mcitedefaultendpunct}{\mcitedefaultseppunct}\relax
\EndOfBibitem
\bibitem[Zhang and Cremer(2010)Zhang, and Cremer]{zhang}
Zhang,~Y.; Cremer,~P. Chemistry of Hofmeister Anions and Osmolytes. \emph{Annu.
  Rev. Phys. Chem.} \textbf{2010}, \emph{61}, 63--83\relax
\mciteBstWouldAddEndPuncttrue
\mciteSetBstMidEndSepPunct{\mcitedefaultmidpunct}
{\mcitedefaultendpunct}{\mcitedefaultseppunct}\relax
\EndOfBibitem
\bibitem[Heyda \latin{et~al.}(2013)Heyda, Muzdalo, and Dzubiella]{Heyda2013}
Heyda,~J.; Muzdalo,~A.; Dzubiella,~J. {Rationalizing polymer swelling and
  collapse under attractive cosolvent conditions}. \emph{Macromolecules}
  \textbf{2013}, \emph{46}, 1231--1238\relax
\mciteBstWouldAddEndPuncttrue
\mciteSetBstMidEndSepPunct{\mcitedefaultmidpunct}
{\mcitedefaultendpunct}{\mcitedefaultseppunct}\relax
\EndOfBibitem
\bibitem[Heyda and Dzubiella(2014)Heyda, and Dzubiella]{Heyda2014}
Heyda,~J.; Dzubiella,~J. {Thermodynamic Description of Hofmeister Effects on
  the LCST of Thermosensitive Polymers}. \emph{J. Phys. Chem. B} \textbf{2014},
  \emph{118}, 10979--10988\relax
\mciteBstWouldAddEndPuncttrue
\mciteSetBstMidEndSepPunct{\mcitedefaultmidpunct}
{\mcitedefaultendpunct}{\mcitedefaultseppunct}\relax
\EndOfBibitem
\bibitem[Zhao \latin{et~al.}(2001)Zhao, Zhang, and Wu]{zhao2001nonergodic}
Zhao,~Y.; Zhang,~G.; Wu,~C. Nonergodic dynamics of a novel thermally sensitive
  hybrid gel. \emph{Macromolecules} \textbf{2001}, \emph{34}, 7804--7808\relax
\mciteBstWouldAddEndPuncttrue
\mciteSetBstMidEndSepPunct{\mcitedefaultmidpunct}
{\mcitedefaultendpunct}{\mcitedefaultseppunct}\relax
\EndOfBibitem
\bibitem[Moncho-Jord{\'a} \latin{et~al.}(2013)Moncho-Jord{\'a}, Anta, and
  Callejas-Fern{\'a}ndez]{moncho2013effective}
Moncho-Jord{\'a},~A.; Anta,~J.; Callejas-Fern{\'a}ndez,~J. Effective
  electrostatic interactions arising in core-shell charged microgel suspensions
  with added salt. \emph{J. Chem. Phys.} \textbf{2013}, \emph{138},
  134902\relax
\mciteBstWouldAddEndPuncttrue
\mciteSetBstMidEndSepPunct{\mcitedefaultmidpunct}
{\mcitedefaultendpunct}{\mcitedefaultseppunct}\relax
\EndOfBibitem
\bibitem[Moncho-Jord{\'a}(2013)]{moncho2013effective2}
Moncho-Jord{\'a},~A. Effective charge of ionic microgel particles in the
  swollen and collapsed states: The role of the steric microgel-ion repulsion.
  \emph{J. Chem. Phys.} \textbf{2013}, \emph{139}, 064906\relax
\mciteBstWouldAddEndPuncttrue
\mciteSetBstMidEndSepPunct{\mcitedefaultmidpunct}
{\mcitedefaultendpunct}{\mcitedefaultseppunct}\relax
\EndOfBibitem
\bibitem[Moncho-Jord{\'a} and Adroher-Ben{\'\i}tez(2014)Moncho-Jord{\'a}, and
  Adroher-Ben{\'\i}tez]{moncho2014ion}
Moncho-Jord{\'a},~A.; Adroher-Ben{\'\i}tez,~I. Ion permeation inside microgel
  particles induced by specific interactions: from charge inversion to
  overcharging. \emph{Soft Matter} \textbf{2014}, \emph{10}, 5810--5823\relax
\mciteBstWouldAddEndPuncttrue
\mciteSetBstMidEndSepPunct{\mcitedefaultmidpunct}
{\mcitedefaultendpunct}{\mcitedefaultseppunct}\relax
\EndOfBibitem
\bibitem[Adroher-Ben{\'\i}tez \latin{et~al.}(2017)Adroher-Ben{\'\i}tez,
  Moncho-Jord{\'a}, and Dzubiella]{doi:10.1021/acs.langmuir.7b00356}
Adroher-Ben{\'\i}tez,~I.; Moncho-Jord{\'a},~A.; Dzubiella,~J. Sorption and
  Spatial Distribution of Protein Globules in Charged Hydrogel Particles.
  \emph{Langmuir} \textbf{2017}, \emph{33}, 4567--4577\relax
\mciteBstWouldAddEndPuncttrue
\mciteSetBstMidEndSepPunct{\mcitedefaultmidpunct}
{\mcitedefaultendpunct}{\mcitedefaultseppunct}\relax
\EndOfBibitem
\bibitem[Kandu\v{c} \latin{et~al.}(2017)Kandu\v{c}, Chudoba, Palczynski, Kim,
  Roa, and Dzubiella]{Matej2017}
Kandu\v{c},~M.; Chudoba,~R.; Palczynski,~K.; Kim,~W.~K.; Roa,~R.; Dzubiella,~J.
  Selective solute adsorption and partitioning around single PNIPAM chains.
  \emph{Phys. Chem. Chem. Phys.} \textbf{2017}, \emph{19}, 5906--5916\relax
\mciteBstWouldAddEndPuncttrue
\mciteSetBstMidEndSepPunct{\mcitedefaultmidpunct}
{\mcitedefaultendpunct}{\mcitedefaultseppunct}\relax
\EndOfBibitem
\bibitem[Tanaka(1978)]{Tanaka1978}
Tanaka,~T. {Collapse of gels and the critical endpoint}. \emph{Phys. Rev.
  Lett.} \textbf{1978}, \emph{40}, 820--823\relax
\mciteBstWouldAddEndPuncttrue
\mciteSetBstMidEndSepPunct{\mcitedefaultmidpunct}
{\mcitedefaultendpunct}{\mcitedefaultseppunct}\relax
\EndOfBibitem
\bibitem[Tanaka \latin{et~al.}(1977)Tanaka, Ishiwata, and
  Ishimoto]{tanaka1977critical}
Tanaka,~T.; Ishiwata,~S.; Ishimoto,~C. Critical Behavior of Density
  Fluctuations in Gels. \emph{Phys. Rev. Lett.} \textbf{1977}, \emph{38},
  771--774\relax
\mciteBstWouldAddEndPuncttrue
\mciteSetBstMidEndSepPunct{\mcitedefaultmidpunct}
{\mcitedefaultendpunct}{\mcitedefaultseppunct}\relax
\EndOfBibitem
\bibitem[Mann \latin{et~al.}(2007)Mann, Everaers, Holm, and Kremer]{Mann2007}
Mann,~B.~A.; Everaers,~R.; Holm,~C.; Kremer,~K. {Scaling in polyelectrolyte
  networks}. \emph{Europhys. Lett.} \textbf{2004}, \emph{67}, 786--792\relax
\mciteBstWouldAddEndPuncttrue
\mciteSetBstMidEndSepPunct{\mcitedefaultmidpunct}
{\mcitedefaultendpunct}{\mcitedefaultseppunct}\relax
\EndOfBibitem
\bibitem[Duering \latin{et~al.}(1994)Duering, Kremer, and Grest]{Duering1994}
Duering,~E.~R.; Kremer,~K.; Grest,~G.~S. Structure and relaxation of end-linked
  polymer networks. \emph{J. Chem. Phys.} \textbf{1994}, \emph{101},
  8169--8192\relax
\mciteBstWouldAddEndPuncttrue
\mciteSetBstMidEndSepPunct{\mcitedefaultmidpunct}
{\mcitedefaultendpunct}{\mcitedefaultseppunct}\relax
\EndOfBibitem
\bibitem[Escobedo and de~Pablo(1997)Escobedo, and de~Pablo]{Escobedo1997}
Escobedo,~F.~A.; de~Pablo,~J.~J. Simulation and theory of the swelling of
  athermal gels. \emph{J. Chem. Phys.} \textbf{1997}, \emph{106},
  793--810\relax
\mciteBstWouldAddEndPuncttrue
\mciteSetBstMidEndSepPunct{\mcitedefaultmidpunct}
{\mcitedefaultendpunct}{\mcitedefaultseppunct}\relax
\EndOfBibitem
\bibitem[Escobedo and De~Pablo(1999)Escobedo, and De~Pablo]{Escobedo1999}
Escobedo,~F.~A.; De~Pablo,~J.~J. Molecular simulation of polymeric networks and
  gels: phase behavior and swelling. \emph{Phys. Rep.} \textbf{1999},
  \emph{318}, 85--112\relax
\mciteBstWouldAddEndPuncttrue
\mciteSetBstMidEndSepPunct{\mcitedefaultmidpunct}
{\mcitedefaultendpunct}{\mcitedefaultseppunct}\relax
\EndOfBibitem
\bibitem[Aydt and Hentschke(2000)Aydt, and Hentschke]{Aydt2000}
Aydt,~E.; Hentschke,~R. Swelling of a model network: A Gibbs-ensemble molecular
  dynamics study. \emph{J. Chem. Phys.} \textbf{2000}, \emph{112},
  5480--5487\relax
\mciteBstWouldAddEndPuncttrue
\mciteSetBstMidEndSepPunct{\mcitedefaultmidpunct}
{\mcitedefaultendpunct}{\mcitedefaultseppunct}\relax
\EndOfBibitem
\bibitem[Lu and Hentschke(2002)Lu, and Hentschke]{Lu2002}
Lu,~Z.~Y.; Hentschke,~R. Swelling of model polymer networks with different
  cross-link densities: A computer simulation study. \emph{Phys. Rev. E}
  \textbf{2002}, \emph{66}, 1--8\relax
\mciteBstWouldAddEndPuncttrue
\mciteSetBstMidEndSepPunct{\mcitedefaultmidpunct}
{\mcitedefaultendpunct}{\mcitedefaultseppunct}\relax
\EndOfBibitem
\bibitem[Schneider and Linse(2002)Schneider, and Linse]{Schneider2002}
Schneider,~S.; Linse,~P. Swelling of cross-linked polyelectrolyte gels.
  \emph{Eur. Phys. J. E} \textbf{2002}, \emph{8}, 457--460\relax
\mciteBstWouldAddEndPuncttrue
\mciteSetBstMidEndSepPunct{\mcitedefaultmidpunct}
{\mcitedefaultendpunct}{\mcitedefaultseppunct}\relax
\EndOfBibitem
\bibitem[Schneider and Linse(2003)Schneider, and Linse]{Schneider2003}
Schneider,~S.; Linse,~P. Monte Carlo Simulation of Defect-Free Cross-Linked
  Polyelectrolyte Gels. \emph{J. Phys. Chem. B} \textbf{2003}, \emph{107},
  8030--8040\relax
\mciteBstWouldAddEndPuncttrue
\mciteSetBstMidEndSepPunct{\mcitedefaultmidpunct}
{\mcitedefaultendpunct}{\mcitedefaultseppunct}\relax
\EndOfBibitem
\bibitem[Schneider and Linse(2004)Schneider, and Linse]{Schneider2004}
Schneider,~S.; Linse,~P. {Discontinuous volume transitions in cross-linked
  polyelectrolyte gels induced by short-range attractions and strong
  electrostatic coupling}. \emph{Macromolecules} \textbf{2004}, \emph{37},
  3850--3856\relax
\mciteBstWouldAddEndPuncttrue
\mciteSetBstMidEndSepPunct{\mcitedefaultmidpunct}
{\mcitedefaultendpunct}{\mcitedefaultseppunct}\relax
\EndOfBibitem
\bibitem[Yan and de~Pablo(2003)Yan, and de~Pablo]{Yan2003}
Yan,~Q.; de~Pablo,~J.~J. {Monte Carlo simulation of a coarse-grained model of
  polyelectrolyte networks.} \emph{Phys. Rev. Lett.} \textbf{2003}, \emph{91},
  018301\relax
\mciteBstWouldAddEndPuncttrue
\mciteSetBstMidEndSepPunct{\mcitedefaultmidpunct}
{\mcitedefaultendpunct}{\mcitedefaultseppunct}\relax
\EndOfBibitem
\bibitem[Edgecombe and Linse(2007)Edgecombe, and Linse]{Edgecombe2007}
Edgecombe,~S.; Linse,~P. {Monte Carlo simulation of polyelectrolyte gels:
  Effects of polydispersity and topological defects}. \emph{Macromolecules}
  \textbf{2007}, \emph{40}, 3868--3875\relax
\mciteBstWouldAddEndPuncttrue
\mciteSetBstMidEndSepPunct{\mcitedefaultmidpunct}
{\mcitedefaultendpunct}{\mcitedefaultseppunct}\relax
\EndOfBibitem
\bibitem[Binder(1995)]{Binder2008}
Binder,~K. \emph{Monte Carlo and molecular dynamics simulations in polymer
  science}; Oxford University Press, 1995\relax
\mciteBstWouldAddEndPuncttrue
\mciteSetBstMidEndSepPunct{\mcitedefaultmidpunct}
{\mcitedefaultendpunct}{\mcitedefaultseppunct}\relax
\EndOfBibitem
\bibitem[Jha \latin{et~al.}(2011)Jha, Zwanikken, Detcheverry, de~Pablo, and
  {Olvera de la Cruz}]{Jha2011}
Jha,~P.; Zwanikken,~J.; Detcheverry,~F.; de~Pablo,~J.; {Olvera de la Cruz},~M.
  {Study of volume phase transitions in polymeric nanogels by theoretically
  informed coarse-grained simulations}. \emph{Soft Matter} \textbf{2011},
  \emph{7}, 5965--5975\relax
\mciteBstWouldAddEndPuncttrue
\mciteSetBstMidEndSepPunct{\mcitedefaultmidpunct}
{\mcitedefaultendpunct}{\mcitedefaultseppunct}\relax
\EndOfBibitem
\bibitem[Jiao and Torquato(2012)Jiao, and Torquato]{Jiao2012}
Jiao,~Y.; Torquato,~S. {Quantitative characterization of the microstructure and
  transport properties of biopolymer networks.} \emph{Phys. Biol.}
  \textbf{2012}, \emph{9}, 036009\relax
\mciteBstWouldAddEndPuncttrue
\mciteSetBstMidEndSepPunct{\mcitedefaultmidpunct}
{\mcitedefaultendpunct}{\mcitedefaultseppunct}\relax
\EndOfBibitem
\bibitem[Quesada-P{\'{e}}rez \latin{et~al.}(2012)Quesada-P{\'{e}}rez, Ramos,
  Forcada, and Mart{\'{i}}n-Molina]{Quesada-Perez2012}
Quesada-P{\'{e}}rez,~M.; Ramos,~J.; Forcada,~J.; Mart{\'{i}}n-Molina,~A.
  {Computer simulations of thermo-sensitive microgels: Quantitative comparison
  with experimental swelling data}. \emph{J. Chem. Phys.} \textbf{2012},
  \emph{136}, 244903\relax
\mciteBstWouldAddEndPuncttrue
\mciteSetBstMidEndSepPunct{\mcitedefaultmidpunct}
{\mcitedefaultendpunct}{\mcitedefaultseppunct}\relax
\EndOfBibitem
\bibitem[Ko{\v{s}}ovan \latin{et~al.}(2013)Ko{\v{s}}ovan, Richter, and
  Holm]{Vagias2013}
Ko{\v{s}}ovan,~P.; Richter,~T.; Holm,~C. \emph{Intelligent Hydrogels};
  Springer, 2013; pp 205--221\relax
\mciteBstWouldAddEndPuncttrue
\mciteSetBstMidEndSepPunct{\mcitedefaultmidpunct}
{\mcitedefaultendpunct}{\mcitedefaultseppunct}\relax
\EndOfBibitem
\bibitem[Ko{\v{s}}ovan \latin{et~al.}(2015)Ko{\v{s}}ovan, Richter, and
  Holm]{Kosovan2015}
Ko{\v{s}}ovan,~P.; Richter,~T.; Holm,~C. {Modeling of Polyelectrolyte Gels in
  Equilibrium with Salt Solutions}. \emph{Macromolecules} \textbf{2015},
  \emph{48}, 7698--7708\relax
\mciteBstWouldAddEndPuncttrue
\mciteSetBstMidEndSepPunct{\mcitedefaultmidpunct}
{\mcitedefaultendpunct}{\mcitedefaultseppunct}\relax
\EndOfBibitem
\bibitem[Kobayashi and Winkler(2016)Kobayashi, and Winkler]{Kobayashi2016}
Kobayashi,~H.; Winkler,~R.~G. {Universal conformational properties of polymers
  in ionic nanogels}. \emph{Sci. Rep.} \textbf{2016}, \emph{6}, 19836\relax
\mciteBstWouldAddEndPuncttrue
\mciteSetBstMidEndSepPunct{\mcitedefaultmidpunct}
{\mcitedefaultendpunct}{\mcitedefaultseppunct}\relax
\EndOfBibitem
\bibitem[Schmid \latin{et~al.}(2016)Schmid, Dubbert, Rudov, Pedersen, Lindner,
  Karg, Potemkin, and Richtering]{Schmid2016}
Schmid,~A.; Dubbert,~J.; Rudov,~A.~A.; Pedersen,~J.; Lindner,~P.; Karg,~M.;
  Potemkin,~I.; Richtering,~W. {Multi-Shell Hollow Nanogels with Responsive
  Shell Permeability}. \emph{Sci. Rep.} \textbf{2016}, 22736\relax
\mciteBstWouldAddEndPuncttrue
\mciteSetBstMidEndSepPunct{\mcitedefaultmidpunct}
{\mcitedefaultendpunct}{\mcitedefaultseppunct}\relax
\EndOfBibitem
\bibitem[Adroher-Ben{\'\i}tez \latin{et~al.}(2015)Adroher-Ben{\'\i}tez,
  Ahualli, Mart{\'\i}n-Molina, Quesada-P{\'e}rez, and
  Moncho-Jord{\'a}]{adroher2015role}
Adroher-Ben{\'\i}tez,~I.; Ahualli,~S.; Mart{\'\i}n-Molina,~A.;
  Quesada-P{\'e}rez,~M.; Moncho-Jord{\'a},~A. Role of steric interactions on
  the ionic permeation inside charged microgels: theory and simulations.
  \emph{Macromolecules} \textbf{2015}, \emph{48}, 4645--4656\relax
\mciteBstWouldAddEndPuncttrue
\mciteSetBstMidEndSepPunct{\mcitedefaultmidpunct}
{\mcitedefaultendpunct}{\mcitedefaultseppunct}\relax
\EndOfBibitem
\bibitem[Adroher-Ben{\'\i}tez \latin{et~al.}(2017)Adroher-Ben{\'\i}tez,
  Mart{\'\i}n-Molina, Ahualli, Quesada-P{\'e}rez, Odriozola, and
  Moncho-Jord{\'a}]{adroher2017competition}
Adroher-Ben{\'\i}tez,~I.; Mart{\'\i}n-Molina,~A.; Ahualli,~S.;
  Quesada-P{\'e}rez,~M.; Odriozola,~G.; Moncho-Jord{\'a},~A. Competition
  between excluded-volume and electrostatic interactions for nanogel swelling:
  effects of the counterion valence and nanogel charge. \emph{Phys. Chem. Chem.
  Phys.} \textbf{2017}, \emph{19}, 6838--6848\relax
\mciteBstWouldAddEndPuncttrue
\mciteSetBstMidEndSepPunct{\mcitedefaultmidpunct}
{\mcitedefaultendpunct}{\mcitedefaultseppunct}\relax
\EndOfBibitem
\bibitem[Netz and Dorfm{\"u}ller(1997)Netz, and Dorfm{\"u}ller]{pnetz1997}
Netz,~P.~A.; Dorfm{\"u}ller,~T. Computer simulation studies of diffusion in
  gels: Model structures. \emph{J. Chem. Phys.} \textbf{1997}, \emph{107},
  9221--9233\relax
\mciteBstWouldAddEndPuncttrue
\mciteSetBstMidEndSepPunct{\mcitedefaultmidpunct}
{\mcitedefaultendpunct}{\mcitedefaultseppunct}\relax
\EndOfBibitem
\bibitem[Erba\c{s} and Olvera de~la Cruz(2015)Erba\c{s}, and Olvera de~la
  Cruz]{Erbas2015}
Erba\c{s},~A.; Olvera de~la Cruz,~M. Energy Conversion in Polyelectrolyte
  Hydrogels. \emph{ACS Macro Lett} \textbf{2015}, \emph{4}, 857--861\relax
\mciteBstWouldAddEndPuncttrue
\mciteSetBstMidEndSepPunct{\mcitedefaultmidpunct}
{\mcitedefaultendpunct}{\mcitedefaultseppunct}\relax
\EndOfBibitem
\bibitem[Erba\c{s} and Olvera de~la Cruz(2016)Erba\c{s}, and Olvera de~la
  Cruz]{Erbas2016}
Erba\c{s},~A.; Olvera de~la Cruz,~M. Interactions between Polyelectrolyte Gel
  Surfaces. \emph{Macromolecules} \textbf{2016}, \emph{49}, 9026--9034\relax
\mciteBstWouldAddEndPuncttrue
\mciteSetBstMidEndSepPunct{\mcitedefaultmidpunct}
{\mcitedefaultendpunct}{\mcitedefaultseppunct}\relax
\EndOfBibitem
\bibitem[Li \latin{et~al.}(2016)Li, Erba\c{s}, Zwanikken, and Olvera de~la
  Cruz]{Li2016}
Li,~H.; Erba\c{s},~A.; Zwanikken,~J.; Olvera de~la Cruz,~M. Ionic Conductivity
  in Polyelectrolyte Hydrogels. \emph{Macromolecules} \textbf{2016}, \emph{49},
  9239--9246\relax
\mciteBstWouldAddEndPuncttrue
\mciteSetBstMidEndSepPunct{\mcitedefaultmidpunct}
{\mcitedefaultendpunct}{\mcitedefaultseppunct}\relax
\EndOfBibitem
\bibitem[Plimpton(1995)]{Plimpton1995}
Plimpton,~S. {Fast Parallel Algorithms for Short-Range Molecular Dynamics}.
  \emph{J. Comput. Phys.} \textbf{1995}, \emph{117}, 1--19\relax
\mciteBstWouldAddEndPuncttrue
\mciteSetBstMidEndSepPunct{\mcitedefaultmidpunct}
{\mcitedefaultendpunct}{\mcitedefaultseppunct}\relax
\EndOfBibitem
\bibitem[Berendsen \latin{et~al.}(1984)Berendsen, Postma, van Gunsteren,
  DiNola, and Haak]{berendsen1984molecular}
Berendsen,~H.~J.; Postma,~J.~v.; van Gunsteren,~W.~F.; DiNola,~A.; Haak,~J.
  Molecular dynamics with coupling to an external bath. \emph{J. Chem. Phys}
  \textbf{1984}, \emph{81}, 3684--3690\relax
\mciteBstWouldAddEndPuncttrue
\mciteSetBstMidEndSepPunct{\mcitedefaultmidpunct}
{\mcitedefaultendpunct}{\mcitedefaultseppunct}\relax
\EndOfBibitem
\bibitem[Kienberger \latin{et~al.}(2000)Kienberger, Pastushenko, Kada, Gruber,
  Riener, Schindler, and Hinterdorfer]{Kienberger2000}
Kienberger,~F.; Pastushenko,~V.~P.; Kada,~G.; Gruber,~H.~J.; Riener,~C.;
  Schindler,~H.; Hinterdorfer,~P. Static and Dynamical Properties of Single
  Poly(Ethylene Glycol) Molecules Investigated by Force Spectroscopy.
  \emph{Single Mol.} \textbf{2000}, \emph{1}, 123--128\relax
\mciteBstWouldAddEndPuncttrue
\mciteSetBstMidEndSepPunct{\mcitedefaultmidpunct}
{\mcitedefaultendpunct}{\mcitedefaultseppunct}\relax
\EndOfBibitem
\bibitem[Lee \latin{et~al.}(2008)Lee, Venable, Mackerell, and Pastor]{Lee2008}
Lee,~H.; Venable,~R.~M.; Mackerell,~A.~D.; Pastor,~R.~W. Molecular dynamics
  studies of polyethylene oxide and polyethylene glycol: hydrodynamic radius
  and shape anisotropy. \emph{Biophys. J.} \textbf{2008}, \emph{95},
  1590--1599\relax
\mciteBstWouldAddEndPuncttrue
\mciteSetBstMidEndSepPunct{\mcitedefaultmidpunct}
{\mcitedefaultendpunct}{\mcitedefaultseppunct}\relax
\EndOfBibitem
\bibitem[Zhang \latin{et~al.}(2000)Zhang, Zou, Wang, and
  Zhang]{zhang2000single}
Zhang,~W.; Zou,~S.; Wang,~C.; Zhang,~X. Single polymer chain elongation of poly
  (N-isopropylacrylamide) and poly (acrylamide) by atomic force microscopy.
  \emph{J. Phys. Chem. B} \textbf{2000}, \emph{104}, 10258--10264\relax
\mciteBstWouldAddEndPuncttrue
\mciteSetBstMidEndSepPunct{\mcitedefaultmidpunct}
{\mcitedefaultendpunct}{\mcitedefaultseppunct}\relax
\EndOfBibitem
\bibitem[Ahmed \latin{et~al.}(2009)Ahmed, Gooding, Pimenov, Wang, and
  Asher]{ahmed2009uv}
Ahmed,~Z.; Gooding,~E.~A.; Pimenov,~K.~V.; Wang,~L.; Asher,~S.~A. UV resonance
  Raman determination of molecular mechanism of poly (N-isopropylacrylamide)
  volume phase transition. \emph{J. Phys. Chem. B} \textbf{2009}, \emph{113},
  4248--4256\relax
\mciteBstWouldAddEndPuncttrue
\mciteSetBstMidEndSepPunct{\mcitedefaultmidpunct}
{\mcitedefaultendpunct}{\mcitedefaultseppunct}\relax
\EndOfBibitem
\bibitem[Kutnyanszky \latin{et~al.}(2012)Kutnyanszky, Embrechts, Hempenius, and
  Vancso]{kutnyanszky2012there}
Kutnyanszky,~E.; Embrechts,~A.; Hempenius,~M.~A.; Vancso,~G.~J. Is there a
  molecular signature of the LCST of single PNIPAM chains as measured by AFM
  force spectroscopy? \emph{Chem. Phys. Lett.} \textbf{2012}, \emph{535},
  126--130\relax
\mciteBstWouldAddEndPuncttrue
\mciteSetBstMidEndSepPunct{\mcitedefaultmidpunct}
{\mcitedefaultendpunct}{\mcitedefaultseppunct}\relax
\EndOfBibitem
\bibitem[Muroga \latin{et~al.}(1985)Muroga, Nada, and Nagasawa]{PMMA}
Muroga,~Y.; Noda,~I.; Nagasawa,~M. Investigation of Local Conformations of
  Polyelectrolytes in Aqueous Solution by Small-Angle X-ray Scattering. 2.
  Local Conformations of Stereoregular Poly(sodium methacrylates).
  \emph{Macromolecules} \textbf{1985}, \emph{18}, 1580--1582\relax
\mciteBstWouldAddEndPuncttrue
\mciteSetBstMidEndSepPunct{\mcitedefaultmidpunct}
{\mcitedefaultendpunct}{\mcitedefaultseppunct}\relax
\EndOfBibitem
\bibitem[Harris and Hearst(1966)Harris, and Hearst]{harris1966polymer}
Harris,~R.; Hearst,~J. On polymer dynamics. \emph{J. Chem. Phys.}
  \textbf{1966}, \emph{44}, 2595--2602\relax
\mciteBstWouldAddEndPuncttrue
\mciteSetBstMidEndSepPunct{\mcitedefaultmidpunct}
{\mcitedefaultendpunct}{\mcitedefaultseppunct}\relax
\EndOfBibitem
\bibitem[Fixman and Kovac(1973)Fixman, and Kovac]{fixman1973polymer}
Fixman,~M.; Kovac,~J. Polymer conformational statistics. III. Modified Gaussian
  models of stiff chains. \emph{J. Chem. Phys.} \textbf{1973}, \emph{58},
  1564--1568\relax
\mciteBstWouldAddEndPuncttrue
\mciteSetBstMidEndSepPunct{\mcitedefaultmidpunct}
{\mcitedefaultendpunct}{\mcitedefaultseppunct}\relax
\EndOfBibitem
\bibitem[Kierfeld \latin{et~al.}(2004)Kierfeld, Niamploy, Sa-yakanit, and
  Lipowsky]{Kierfeld2004}
Kierfeld,~J.; Niamploy,~O.; Sa-yakanit,~V.; Lipowsky,~R. Stretching of
  semiflexible polymers with elastic bonds. \emph{Eur. Phys. J. E}
  \textbf{2004}, \emph{14}, 17--34\relax
\mciteBstWouldAddEndPuncttrue
\mciteSetBstMidEndSepPunct{\mcitedefaultmidpunct}
{\mcitedefaultendpunct}{\mcitedefaultseppunct}\relax
\EndOfBibitem
\bibitem[Rubinstein and Colby(2003)Rubinstein, and Colby]{Rubinstein}
Rubinstein,~M.; Colby,~R.~H. \emph{Polymer Physics}; Oxford University Press,
  2003\relax
\mciteBstWouldAddEndPuncttrue
\mciteSetBstMidEndSepPunct{\mcitedefaultmidpunct}
{\mcitedefaultendpunct}{\mcitedefaultseppunct}\relax
\EndOfBibitem
\bibitem[Blundell and Terentjev(2009)Blundell, and Terentjev]{Blundell2009}
Blundell,~J.~R.; Terentjev,~E.~M. Stretching Semiflexible Filaments and Their
  Networks. \emph{Macromolecules} \textbf{2009}, \emph{42}, 5388--5394\relax
\mciteBstWouldAddEndPuncttrue
\mciteSetBstMidEndSepPunct{\mcitedefaultmidpunct}
{\mcitedefaultendpunct}{\mcitedefaultseppunct}\relax
\EndOfBibitem
\bibitem[Meng and Terentjev(2017)Meng, and Terentjev]{Meng2017}
Meng,~F.; Terentjev,~E.~M. Theory of Semiflexible Filaments and Networks.
  \emph{Polymers} \textbf{2017}, \emph{9}, 52\relax
\mciteBstWouldAddEndPuncttrue
\mciteSetBstMidEndSepPunct{\mcitedefaultmidpunct}
{\mcitedefaultendpunct}{\mcitedefaultseppunct}\relax
\EndOfBibitem
\bibitem[Edwards and Freed(1969)Edwards, and Freed]{Edwards1969}
Edwards,~S.~F.; Freed,~K.~F. The entropy of a confined polymer: I. \emph{J.
  Phys. A} \textbf{1969}, \emph{2}, 145--150\relax
\mciteBstWouldAddEndPuncttrue
\mciteSetBstMidEndSepPunct{\mcitedefaultmidpunct}
{\mcitedefaultendpunct}{\mcitedefaultseppunct}\relax
\EndOfBibitem
\bibitem[Grosberg and Kuznetsov(1992)Grosberg, and
  Kuznetsov]{grosberg1992quantitative}
Grosberg,~A.~Y.; Kuznetsov,~D. Quantitative theory of the globule-to-coil
  transition. 1. Link density distribution in a globule and its radius of
  gyration. \emph{Macromolecules} \textbf{1992}, \emph{25}, 1970--1979\relax
\mciteBstWouldAddEndPuncttrue
\mciteSetBstMidEndSepPunct{\mcitedefaultmidpunct}
{\mcitedefaultendpunct}{\mcitedefaultseppunct}\relax
\EndOfBibitem
\bibitem[Li and Tanaka(1992)Li, and Tanaka]{Tanaka1992}
Li,~Y.; Tanaka,~T. Phase transitions of gels. \emph{Annu. Rev. Mater. Sci.}
  \textbf{1992}, \emph{22}, 243--277\relax
\mciteBstWouldAddEndPuncttrue
\mciteSetBstMidEndSepPunct{\mcitedefaultmidpunct}
{\mcitedefaultendpunct}{\mcitedefaultseppunct}\relax
\EndOfBibitem
\bibitem[Mukherji and Kremer(2013)Mukherji, and Kremer]{mukherji2013coil}
Mukherji,~D.; Kremer,~K. Coil--globule--coil transition of pnipam in aqueous
  methanol: Coupling all-atom simulations to semi-grand canonical
  coarse-grained reservoir. \emph{Macromolecules} \textbf{2013}, \emph{46},
  9158--9163\relax
\mciteBstWouldAddEndPuncttrue
\mciteSetBstMidEndSepPunct{\mcitedefaultmidpunct}
{\mcitedefaultendpunct}{\mcitedefaultseppunct}\relax
\EndOfBibitem
\bibitem[Mukherji \latin{et~al.}(2014)Mukherji, Marques, and
  Kremer]{mukherji2014polymer}
Mukherji,~D.; Marques,~C.~M.; Kremer,~K. Polymer collapse in miscible good
  solvents is a generic phenomenon driven by preferential adsorption.
  \emph{Nat. Commun.} \textbf{2014}, \emph{5}, 4882\relax
\mciteBstWouldAddEndPuncttrue
\mciteSetBstMidEndSepPunct{\mcitedefaultmidpunct}
{\mcitedefaultendpunct}{\mcitedefaultseppunct}\relax
\EndOfBibitem
\bibitem[Rodr{\'\i}guez-Ropero \latin{et~al.}(2015)Rodr{\'\i}guez-Ropero,
  Hajari, and van~der Vegt]{rodriguez2015mechanism}
Rodr{\'\i}guez-Ropero,~F.; Hajari,~T.; van~der Vegt,~N.~F. Mechanism of polymer
  collapse in miscible good solvents. \emph{J. Phys. Chem. B} \textbf{2015},
  \emph{119}, 15780--15788\relax
\mciteBstWouldAddEndPuncttrue
\mciteSetBstMidEndSepPunct{\mcitedefaultmidpunct}
{\mcitedefaultendpunct}{\mcitedefaultseppunct}\relax
\EndOfBibitem
\bibitem[Rodr{\'\i}guez-Ropero and van~der Vegt(2015)Rodr{\'\i}guez-Ropero, and
  van~der Vegt]{rodriguez2015urea}
Rodr{\'\i}guez-Ropero,~F.; van~der Vegt,~N.~F. On the urea induced hydrophobic
  collapse of a water soluble polymer. \emph{Phys. Chem. Chem. Phys.}
  \textbf{2015}, \emph{17}, 8491--8498\relax
\mciteBstWouldAddEndPuncttrue
\mciteSetBstMidEndSepPunct{\mcitedefaultmidpunct}
{\mcitedefaultendpunct}{\mcitedefaultseppunct}\relax
\EndOfBibitem
\bibitem[Rika \latin{et~al.}(1990)Rika, Meewes, Nyffenegger, and
  Binkert]{rika1990intermolecular}
Rika,~J.; Meewes,~M.; Nyffenegger,~R.; Binkert,~T. Intermolecular and
  intramolecular solubilization: Collapse and expansion of a polymer chain in
  surfactant solutions. \emph{Phys. Rev. Lett.} \textbf{1990}, \emph{65},
  657\relax
\mciteBstWouldAddEndPuncttrue
\mciteSetBstMidEndSepPunct{\mcitedefaultmidpunct}
{\mcitedefaultendpunct}{\mcitedefaultseppunct}\relax
\EndOfBibitem
\bibitem[Lee and Cabane(1997)Lee, and Cabane]{lee1997effects}
Lee,~L.-T.; Cabane,~B. Effects of surfactants on thermally collapsed poly
  (N-isopropylacrylamide) macromolecules. \emph{Macromolecules} \textbf{1997},
  \emph{30}, 6559--6566\relax
\mciteBstWouldAddEndPuncttrue
\mciteSetBstMidEndSepPunct{\mcitedefaultmidpunct}
{\mcitedefaultendpunct}{\mcitedefaultseppunct}\relax
\EndOfBibitem
\bibitem[Heyda \latin{et~al.}(2017)Heyda, Okur, Hlad\'{i}lkov\'{a}, Rembert,
  Hunn, Yang, Dzubiella, Jungwirth, and Cremer]{heyda2017guanidinium}
Heyda,~J.; Okur,~H.~I.; Hlad\'{i}lkov\'{a},~J.; Rembert,~K.~B.; Hunn,~W.;
  Yang,~T.; Dzubiella,~J.; Jungwirth,~P.; Cremer,~P.~S. Guanidinium can both
  Cause and Prevent the Hydrophobic Collapse of Biomacromolecules. \emph{J. Am.
  Chem. Soc.} \textbf{2017}, \emph{139}, 863--870\relax
\mciteBstWouldAddEndPuncttrue
\mciteSetBstMidEndSepPunct{\mcitedefaultmidpunct}
{\mcitedefaultendpunct}{\mcitedefaultseppunct}\relax
\EndOfBibitem
\bibitem[Moncho-Jord{\'a} and Dzubiella(2016)Moncho-Jord{\'a}, and
  Dzubiella]{moncho2016swelling}
Moncho-Jord{\'a},~A.; Dzubiella,~J. Swelling of ionic microgel particles in the
  presence of excluded-volume interactions: a density functional approach.
  \emph{Phys. Chem. Chem. Phys.} \textbf{2016}, \emph{18}, 5372--5385\relax
\mciteBstWouldAddEndPuncttrue
\mciteSetBstMidEndSepPunct{\mcitedefaultmidpunct}
{\mcitedefaultendpunct}{\mcitedefaultseppunct}\relax
\EndOfBibitem
\bibitem[Hansing \latin{et~al.}(2016)Hansing, Ciemer, Kim, Zhang, DeRouchey,
  and Netz]{Johan2016}
Hansing,~J.; Ciemer,~C.; Kim,~W.~K.; Zhang,~X.; DeRouchey,~J.~E.; Netz,~R.~R.
  Nanoparticle filtering in charged hydrogels: Effects of particle size, charge
  asymmetry and salt concentration. \emph{Eur. Phys. J. E} \textbf{2016},
  \emph{39}, 1--13\relax
\mciteBstWouldAddEndPuncttrue
\mciteSetBstMidEndSepPunct{\mcitedefaultmidpunct}
{\mcitedefaultendpunct}{\mcitedefaultseppunct}\relax
\EndOfBibitem
\end{mcitethebibliography}


\providecommand{\latin}[1]{#1}
\providecommand*\mcitethebibliography{\thebibliography}
\csname @ifundefined\endcsname{endmcitethebibliography}
  {\let\endmcitethebibliography\endthebibliography}{}


%

\end{document}